\documentclass[onecolumn]{emulateapj-rtx4}

\usepackage{amsmath}
\usepackage{color}    %

\newcommand{\BF}[1]{\mbox{\normalsize \boldmath $#1$}}
\newcommand{\pfrac}[2]{\frac{\partial #1}{\partial #2}}
\newcommand{\kakkoi}[3]{\left(\frac{{#1}}{{#2}}\right)^{#3}}

\renewcommand\thefootnote{{\color{red}*\arabic{footnote}}}

\slugcomment{Accepted for publication in ApJ}

\shorttitle{Accretion onto Circumplanetary Disks}
\shortauthors{Tanigawa et al.}

\begin{document}

\title{Distribution of Accreting Gas and Angular Momentum onto
    Circumplanetary Disks}


\author{Takayuki Tanigawa}
\affil{Center for Planetary Science; Institute of Low Temperature
Science, Hokkaido University, Sapporo, Japan}
\email{tanigawa@cps-jp.org}

\author{Keiji Ohtsuki}
\affil{Department of Earth and Planetary Sciences, Kobe University,
Kobe, Japan; and Laboratory for Atmospheric and Space Physics,
University of Colorado, Boulder, Colorado}

\and

\author{Masahiro N. Machida} \affil{Department of Earth and Planetary
Sciences, Graduate School of Sciences, Kyushu University, Fukuoka, Japan}




\begin{abstract}
We investigate gas accretion flow onto a circumplanetary disk from a
protoplanetary disk in detail by using high-resolution three-dimensional
nested-grid hydrodynamic simulations, in order to provide a basis of
formation processes of satellites around giant planets.
Based on detailed analyses of gas accretion flow, we find that most of
gas accretion onto circumplanetary disks occurs nearly vertically toward
the disk surface from high altitude,
which generates a shock surface at several scale heights of the
circumplanetary disk.
The gas that has passed through the shock surface moves inward because
its specific angular momentum is smaller than that of the local
Keplerian rotation, while gas near the midplane in the protoplanetary
disk cannot accrete to the circumplanetary disk.
Gas near the midplane within the planet's Hill sphere spirals outward
and escapes from the Hill sphere through the two Lagrangian points L$_1$
and L$_2$.
We also analyze fluxes of accreting mass and angular momentum in detail
and find that the distributions of the fluxes onto the disk surface are
well described by power-law functions and that a large fraction of gas
accretion occurs at the outer region of the disk, i.e., at about 0.1
times the Hill radius.  The nature of power-law functions indicates
that, other than the outer edge, there is no specific radius where gas
accretion is concentrated.
These source functions of mass and angular momentum in the
circumplanetary disk would provide us with useful constraints on the
structure and evolution of the circumplanetary disk, which is important
for satellite formation.
\end{abstract}






\section{Introduction}\label{sec_introduction}
Satellite systems around the giant planets in our solar system are
commonly seen.  They are thought to have formed in circumplanetary
disks, which are believed to have existed around giant planets during
their gas capturing growing stage.

In earlier works, formation process of satellite systems have been
considered based on a minimum mass subnebula (MMSN) model, in which
satellites form from a disk that contains sufficient solid mass with
solar composition for reproducing the current satellite systems (e.g.,
\citet{Lunine82}), as an analog of the minimum mass {\it solar} nebula
model \citep{Hayashi81}.  However, it was suggested that the MMSN model
has difficulty in reproducing current satellite systems around Jupiter
and Saturn \citep{Canup02}.  One of severe problems is that the model
leads to much higher temperature than that of H$_2$O ice sublimation at
the current regular satellite region, which means that ice, which is the
main component of the satellites, cannot be used as building material of
the satellites.

In order to overcome the difficulties of the MMSN-type models which
assumes a closed and static disk, alternative models have been
developed.
\citet{Canup02} proposed a model in which an accretion disk with
continuous supply of gas and solid is considered as a proto-satellite
disk.
This model is based on results of hydrodynamic simulations of gas
capturing process of giant planets \citep[e.g.,][]{Lubow99, DAngelo02},
which demonstrated that gas accretion from the protoplanetary disk
toward the parent planets occurs through a
circumplanetary disk.
In this model, the surface density and temperature of the
circumplanetary disk is kept lower than assumed in the MMSN model, and
thus H$_2$O ice can be used as a solid building material of satellites.
Based on comparison among time scales of different processes, they
concluded that formation of the Galilean satellites can be best
explained if the circumplanetary disk had one order of magnitude lower
gas surface density than the MMSN disk, with slow gas accretion rate
corresponding to Jupiter's growth time scale longer than a few $\times
10^6$ yr (see also \citet{Sasaki10}).
On the other hand, \citet{Mosqueira03a} proposed another disk model
which consists of two components, i.e., inner MMSN-type massive disk and
outer low-density extended disk.  The model reproduced, for example,
three inner Galilean satellites as well as only partially differentiated
Callisto.
Although these models seem to reproduce the current satellite systems of
the giant planets in our solar system, they needed to assume parameters
for the disk structure such as surface density profile as a basis of
physical processes of satellite formation.

%
The structure of a circumplanetary disk is closely related to the gas
accretion process of giant planets.  Gas flow around proto-giant-planets
in protoplanetary disks has been studied using hydrodynamic simulations.
%
Earlier two-dimensional simulations showed that the gas in the Hill
sphere rotates in the prograde direction \citep{Miki82, Sekiya87,
Korycansky96} and a pair of spiral shocks stands in the circumplanetary
disks \citep[e.g.,][]{Kley99, Lubow99, Tanigawa02, DAngelo02}.  However,
the scale height of the proto-planetary disk for Jupiter-sized planets
is comparable to the Hill radius, thus simulations with two-dimension
approximation cannot capture the feature of the accretion flow.
%
Recent three-dimensional hydrodynamic simulations revealed that the
two-dimensional picture for circumplanetary disks is not appropriate
for the flow in the Hill sphere
\citep[e.g.,][]{DAngelo03,Bate03,Machida08,Paardekooper08,Ayliffe09b,Coradini10}.
One of the most important features newly found in three-dimensional
calculations is that, in the region of circumplanetary disks, the gas
accretes nearly vertically downward toward the midplane from high
altitude.

Although some studies using hydrodynamic simulations were able to obtain
the structures of circumplanetary disks, the direct use of results of
hydrodynamic simulations for the structure, such as surface density,
is problematic.
One of the reasons is that disk-like shear-dominant flow is susceptible
to numerical viscosity, which is intrinsic nature of numerical
hydrodynamic calculation, no matter if SPH method or mesh based method
is used, thus it is difficult to obtain reliable structure of the disk
directly from such simulations.
In particular, orbital radius of the rotating gas in disks basically
changes only slightly with time, thus numerical error tends to
accumulate easily in simulation of long-term evolution.
Also timescale required for low viscous disks such as circumplanetary
disks to reach steady state is much longer than the typical dynamical
time of the fluid.  While high-resolution hydrodynamic simulations is
needed to resolve the structure of circumplanetary disks, their
long-term evolution is difficult to follow with such time-consuming simulation.
In addition, physical (non-numerical) viscosity in the disk is not well
understood, and we need to assume specific viscosity models that are
hard to justify.

%
As an alternative approach, in the present work, we examine gas
accretion flow onto circumplanetary disks from proto-planetary disks,
in order to determine gas accretion rate as a function of distance from
parent planets.
Unlike the structure in a rotation disk, the accretion flow onto
circumplanetary disks is not susceptible to numerical viscosity.
Also, because the circumplanetary disks are located at the downstream of
super-sonic accretion flow,
the accretion flow itself is hardly affected by the circumplanetary
disk structure, which depends on poorly-known effective viscosity.
In \S \ref{sec_methods}, we describe our settings of hydrodynamic
simulations.  In \S \ref{sec_results}, results and analyses of the
simulations are shown.  We discuss implication of our analyses of the
accretion flow in \S \ref{sec_discussion}.  We summarize our results in
\S \ref{sec_summary}.

\section{Methods}\label{sec_methods}
\subsection{Settings and Basic Equations}
We consider a situation in which a growing giant planet embedded in a
protoplanetary disk has induced the nucleated instability and the gas of
the disk accretes dynamically onto the planet.  We take local Cartesian
coordinates rotating with the planet, $x$-axis corresponding to the
radial direction, $y$-axis the revolving direction of the planet, and
$z$-axis normal to the disk midplane.  The planet is located at the
origin.

We adopt the local approximation, in which tidal potential is linearized
and curvature of the protoplanetary disk is neglected.  This
approximation is valid as long as the Hill radius is much smaller than
the orbital radius of the planet.
The orbit of the planet is assumed to be fixed circular and co-planar
with the disk midplane.  The gas is assumed to be inviscid and
isothermal.  Magnetic field and self-gravity of the gas are neglected.
%

We employ equation of continuity, equation of motion for compressive
inviscid gas, and equation of state for isothermal gas to simulate the
gas motion.  In order to understand physics clearly, we use these
equations in a non-dimensional form.  We normalize time by the inverse
of the Keplerian angular velocity $\Omega_{\rm K}^{-1} \equiv
(GM_\ast/a^3)^{-1/2}$, length by scale height $h\equiv c\Omega_{\rm
K}^{-1}$, and mass by unperturbed surface density of the protoplanetary
disk gas $\Sigma_0$ times $h^2$, where $G$ is the gravitational
constant, $M_\ast$ is the solar mass, $a$ is the semi-major axis of the
planet, and $c$ is the sound speed of the disk gas on the planet orbit.
The sound speed $c$ is unity in our normalization.  Normalized
quantities are denoted with tilde on the variables in this paper.  The
normalized equations can be written as
\begin{equation}
\pfrac{\tilde{\rho}}{\tilde{t}}
 + \tilde{\nabla} \cdot (\tilde{\rho} \tilde{\BF{v}})
 = 0,
\end{equation}
\begin{equation}
\pfrac{\BF{v}}{\tilde{t}}
 + (\tilde{\BF{v}} \cdot \tilde{\nabla}) \tilde{\BF{v}}
 = - \frac{1}{\tilde{\rho}} \tilde{\nabla} \tilde{P}
   - \tilde{\nabla} \tilde{\Phi}
   - 2\BF{e}_z \times \tilde{\BF{v}},
\end{equation}
\begin{equation}
\tilde{P} = \tilde{\rho},
\end{equation}
\begin{equation}
\tilde{\Phi}
 = \tilde{\Phi}_{\rm tidal}
  +\tilde{\Phi}_{\rm p}
  + \frac{9}{2} \tilde{r}_{\rm H}^2,
\label{Phi}
\end{equation}
where
\begin{equation}
\tilde{\Phi}_{\rm tidal}
 = -\frac{3}{2} \tilde{x}^2
   +\frac{1}{2} \tilde{z}^2,
\end{equation}
\begin{equation}
\tilde{\Phi}_{\rm p}
 = -\frac{3\tilde{r}_{\rm H}^3}{\tilde{r}}.
\end{equation}
In the above $\tilde{r} = |\tilde{\BF{r}}|$ is distance from the origin,
$\tilde{r}_{\rm H}$ is normalized Hill radius, $\tilde{r}_{\rm H} \equiv
r_{\rm H}/h = (M_{\rm p}/3M_\ast)^{1/3} (a/h)$, $M_{\rm p}$ is the
planet mass, and $\BF{e}_z \equiv (0,0,1)$ is a unit vector in the
$z$-direction.  The third term of the right hand side in Eq.~(\ref{Phi})
is added so that $\tilde{\Phi}=0$ at the two Lagrangian points L$_1$ and
L$_2$ (i.e., $(\tilde{x},\tilde{y},\tilde{z}) = (\pm \tilde{r}_{\rm
H},0,0)$).  The normalized quantities can be written, for example, as
$\tilde{\BF{v}} = \BF{v}/(h\Omega_{\rm K}) = \BF{v}/c$, $\tilde{\rho} =
\rho/(\Sigma_0/(\sqrt{2\pi} h))$, $\tilde{P} = P/(c^2
(\Sigma_0/(\sqrt{2\pi} h)))$, and $\tilde{\Phi} = \Phi/c^2$ for
velocity, density, pressure, and potential, respectively.  Full equation
set before the normalization can be found in \citet{Machida08}.  In the
normalized system, $\tilde{r}_{\rm H}$ is the only characteristic
physical parameter of the system, if we do not consider physical radius
of the planet.  In this paper, we focus on the case with $\tilde{r}_{\rm
H} = 1$, which roughly corresponds to Jupiter mass at 5AU.

\subsection{Numerical Modeling}
We employ a three-dimensional nested-grid hydrodynamic simulation code
\citep[e.g.,][]{MachidaMTH05, MachidaMHT06}, which was originally
developed to explore star formation process by collapse of molecular
cloud core \citep{Matsumoto03b}.
Size of the whole computational domain $(\tilde{L}_x, \tilde{L}_y,
\tilde{L}_z)$ is (24,24,6).  The domain has a symmetry about
$\tilde{z}=0$ plane, and the covered region in simulation is $\tilde{x}
= [-12,12]$, $\tilde{y} = [-12,12]$, and $\tilde{z} = [0,6]$.
We set 11 levels for the nested-grid system, and the numbers of grids in
each level are $(n_x,n_y,n_z) = (64,64,16)$.
The level of the nested grid is denoted by $l$ and $l=1$ is the largest
grid level.  Increment of $l$ by one reduces the size of the
computational domain to half in all directions, keeping its center at
the planet.
The finest grid size is thus $\tilde{L}_x/64/2^{11-1} = 0.000366$, which
corresponds to about 1/4 of the present Jovian radius at 5.2 AU.

%
We adopt two types of artificial gravitational weakening, which allows
us to avoid non-physical crash of the computation without essentially
changing results.
One is the widely-used weakening for region adjacent to the planet to
avoid singularity of the planet gravity.  The modified potential is
given by
\begin{equation}
\tilde{\Phi}_{\rm p}
 = -\frac{3\tilde{r}_{\rm H}^3}{(\tilde{r}^2+\tilde{r}_{\rm sm}^2)^{1/2}},
\end{equation}
where $\tilde{r}_{\rm sm}$ is so-called smoothing length and we set
$\tilde{r}_{\rm sm} = 0.00073$, which corresponds to two grid size of
the finest grid level ($l=11$).
The other gravitational weakening is also applied for the $z$-component
of the tidal force in the high-$z$ region.  In this region, tidal force
in the $z$-direction (toward the mid-plane) is very strong, and causes
too large density gradient to be described by the grid size we set in
the code.  Thus we connect two parabolic curves smoothly at $\tilde{z} =
\tilde{L}_z^\ast$ as
\begin{eqnarray}
\tilde{\Phi}_{{\rm tidal},z}
 = \left\{
\begin{array}{ll}
 \displaystyle
 \frac{1}{2} \tilde{z}^2
    & \mbox{if $\tilde{z} \leq \tilde{L}_z^\ast$}, \vspace{1mm} \\
 \displaystyle
 \frac{1}{2}
 \left\{ -\frac{\tilde{L}_z^\ast}{\tilde{L}_z-\tilde{L}_z^\ast}
          (\tilde{L}_z - \tilde{z})^2
         +\tilde{L}_z\tilde{L}_z^\ast
 \right\}
    & \mbox{if $\tilde{L}_z^\ast \leq \tilde{z} \leq \tilde{L}_z$},
\end{array}
\right.
\label{Phi_tidal_z}
\end{eqnarray}
where we set $\tilde{L}_z^\ast = 4$.  Initial density profile in the
$z$-direction is modified accordingly as
\begin{eqnarray}
\tilde{\rho}(\tilde{x},\tilde{z})
 = \left\{
\begin{array}{ll}
 \displaystyle
 \frac{\tilde{\Sigma}_{\rm ini}(\tilde{x})}{\sqrt{2\pi}}
 \exp\left[ -\frac{\tilde{z}^2}{2} \right]
   & \mbox{if $\tilde{z} \leq \tilde{L}_z^\ast$}, \vspace{2mm} \\
 \displaystyle
 \frac{\tilde{\Sigma}_{\rm ini}(\tilde{x})}{\sqrt{2\pi}}
 \exp\left[ -\frac{1}{2}
             \left\{
                    -\frac{\tilde{L}_z^\ast}{\tilde{L}_z-\tilde{L}_z^\ast}
                     (\tilde{L}_z-\tilde{z})^2
                    +\tilde{L}_z \tilde{L}_z^\ast
             \right\}
     \right]
   & \mbox{if $\tilde{L}_z^\ast \leq \tilde{z} \leq \tilde{L}_z$}, \\
\end{array}
\right.
\end{eqnarray}
for hydrostatic equilibrium under the potential given by
Eq.(\ref{Phi_tidal_z}).  This weakening does not cause any significant
effect for the region where we need to observe because the mass above
$\tilde{L}^\ast$ is negligible (0.0063 \% of the total mass for the
hydrostatic structure).

As boundary conditions, we set the flow at $\tilde{x} = \pm
\tilde{L}_x/2$ to be the unperturbed one, and the mirror condition is
applied at $\tilde{z}=0$ and $\tilde{L}_z$.  As for the boundaries at
$\tilde{y} = \pm \tilde{L}_y/2$ and $\tilde{x} \gtrless 0$ (inflow
region), we use mixed boundary conditions as follows.  When
$\tilde{t}=0$, we set unperturbed condition, which is the same as the
initial condition, and change it gradually to periodic boundary
condition until $\tilde{t} = 100$ based on linear interpolation with
respect to time.  When $\tilde{t} \geq 100$, periodic boundary condition
is applied to all region at $\tilde{y}= \pm \tilde{L}_y/2$.
For the initial condition, we adopt unperturbed density and velocity;
$\tilde{v}_x = \tilde{v}_z = 0$, $\tilde{v}_y = -(3/2)\tilde{x}$ for
velocity and $\partial \tilde{\rho}/\partial \tilde{x} = \partial
\tilde{\rho}/\partial \tilde{y} = 0$, $\tilde{\rho}(\tilde{z}) =
(2\pi)^{-1/2} \exp(-\tilde{z}^2/2)$ for density, which is identical to
that of \citet{Machida08}.

In order to obtain high-resolution results efficiently, calculations are
started with only the largest grid level and the number of levels is
increased with time.  The second largest level ($l=2$) starts at
$\tilde{t}=20$, $l=3$ at $\tilde{t}=50$, and the higher levels ($l \geq
4$) are started in order from $\tilde{t}=150$ by considering relaxation
degree in each level.
In this paper, we will show a snapshot of the flow of $\tilde{t} =
160.7$, when the accretion flow is nearly in the steady state.

We set sink cells around the origin in order to see the pure effect of
gas accretion flow without the effect of the planet body, as well as to
mimic gas accretion phase onto the planet that follows the nucleated
instability \citep[e.g.,][]{Mizuno80, Bodenheimer86, Ikoma00}.  In the
sink cells, gas is removed at a rate that corresponds to
$\tilde{\rho}/\dot{\tilde{\rho}} = 10^{-4}$.  We set $\tilde{r}_{\rm
sink} = \tilde{r}_{\rm sm}$.

\section{Results} \label{sec_results}

\subsection{Path to Circumplanetary Disks}\label{sec_path}
First, we examine the overall structure of accreting gas flow onto a
circumplanetary disk.  Figure \ref{Fig_streamlines} shows streamlines
from outside of the Hill sphere (i.e., from a protoplanetary disk)
toward the planet.  The starting point of a streamline is defined by
$(\tilde{x}_0, \tilde{y}_0, \tilde{z}_0)$.  Four panels show streamlines
starting from four different height; $\tilde{z}_0 =$ 0.0, 0.5, 1.0, 1.5.
Based on the destination of streamlines, we divide the flow into three
regions:
{\it passing region} where gas approaches the planet and passes by
it without making U-turn,
{\it horseshoe region} where gas crosses the planet orbit with U-turn
during the encounter,
and {\it accreting region} where gas is accreted onto the
circumplanetary disk.
We find that there are no accreting streamlines on the midplane
($\tilde{z}=0$), while some streamlines starting from off-midplane
region accretes to the circumplanetary disk.  This implies that the gas
near the midplane in proto-planetary disks is harder to accrete to
circumplanetary disks and planets.
This is confirmed by Figure \ref{Fig_streamlines_outward}, which shows
streamlines starting from three different radii $\tilde{R} \equiv
\sqrt{\tilde{x}^2+\tilde{y}^2} =$ 0.05, 0.1, 0.2, that is,
$(\tilde{x_0}, \tilde{y}_0, \tilde{z}_0) = (0.05,0,0), (0.1,0,0),
(0.2,0,0)$.  We can see that gas on the midplane in $0.2 \lesssim
\tilde{R} \lesssim 1$ spirals outward and escapes from the Hill sphere
within a short timescale through one of the two Lagrangian points (L$_1$
or L$_2$), and that the outward radial velocity decreases with
decreasing distance from the planet (see also Fig.~\ref{Fig_vr-per-vKep}).

The vertical structure of the flow is more clearly demonstrated in
Figure \ref{Fig_fate}.  In this figure, initial positions of streamlines
on the $\tilde{x}$--$\tilde{z}$ plane at $\tilde{y}=\tilde{L}_y/2$ are
classified into three regions by their destinations.  We define the
passing region where streamlines reach the boundary at
$\tilde{y}=-\tilde{L}_y/2$ with $\tilde{x}>0$, horseshoe region where
streamlines reach the boundary at $\tilde{y}=\tilde{L}_y/2$ with
$\tilde{x}<0$, and accreting region where streamlines end up within a
sphere with radius $\tilde{r}_{\rm b}$; we set $\tilde{r}_{\rm b} =
0.2$.  We can clearly see that there is no accretion band in the
midplane, while off-midplane site has an accretion band with a
significant width.  Note that the classification of the three regions
does not depend on $\tilde{r}_{\rm b}$ as long as $0.1 \lesssim
\tilde{r}_{\rm b} \lesssim \tilde{L}_y/2$.  This implies that, even in
the Hill sphere, the gas in the region $\tilde{r} \gtrsim 0.1$ stays
there temporarily and is not necessarily captured by the planet.  We
will discuss the flow in this region in \S \ref{sec_disk_size}.

%

\subsection{Disk Structure and Gas Motion}\label{sec_diskstructure}
%
Figure \ref{Fig_diskstructure} shows azimuthally averaged density,
velocity, and specific angular momentum of the gas with three different
scales.
We first recognize that the gas clearly forms a disk-like structure; the
density distribution is concentrated near the midplane, almost Keplerian
rotation is realized in the high density region (judging from the
contour lines of specific angular momentum), and the flow velocity in
the $R$--$z$ direction of the region is very low.
The density profile in the $z$-direction can be roughly described by
hydrostatic equilibrium in the region where $\tilde{z} \lesssim 5
\tilde{h}_{\rm p}$ and $\tilde{R} \lesssim 0.2$, where $\tilde{h}_{\rm
p}$ is scale height of the circumplanetary disk defined by
\begin{equation}
\tilde{h}_{\rm p}
 \equiv \sqrt{\tilde{R}^3/(3\tilde{r}_{\rm H}^3)}
 \propto \tilde{R}^{3/2}.
\label{h_p}
\end{equation}
From this dependence a flare-up disk is
expected, which is seen indeed in Fig.~\ref{Fig_diskstructure}.
Above the disk surface, large downward velocity, which is almost free
fall velocity, is observed; this shows that gas is indeed accreted
directly onto the disk surface.  The accreting gas forms a shock
surface, which stands at $\tilde{z} \sim 5\tilde{h}_{\rm p}$.
By analyzing streamlines, we confirmed that these gas elements are
actually originated from off-midplane gas (mostly from $\tilde{z}>0.5$)
in the protoplanetary disk.
Also, the contour lines above the shock surface are aligned with the
velocity vectors, which means that specific angular momentum does not
change very much before the gas elements hit the disk surface in this
inner region, where three-body effect is weak enough to be neglected.

Figure \ref{Fig_flux_at_spheres} shows radial mass flux $\tilde{\rho}
\tilde{v}_r$ (where $\tilde{v}_r \equiv \tilde{\BF{v}} \cdot
\tilde{\BF{r}}/\tilde{r}$), as a function of the azimuth and elevation
angles at spheres with four different radii $\tilde{r} = 1.0$, 0.3, 0.1,
0.03.
This figure clearly shows that the accretion manner strongly depends on
both these angles and is not spherically symmetric at all.
For example, the flux at the $\tilde{r}=1.0$ sphere shows that the mass
flux can be both inward and outward directions near the midplane (where
$|\theta| \lesssim 40^\circ$, where $\theta$ is elevation angle)
depending on azimuth angle $\phi$, while it is always inward (i.e.,
negative flux) at high $|\theta|$.
The two maxima near $\phi = 0$ and 180 on the midplane ($\theta = 0$) in
the case of $\tilde{r}=1.0$ corresponds to the outflow from the Hill
sphere through the two Lagrangian points L$_1$ and L$_2$ shown in
Fig.~\ref{Fig_streamlines_outward}.
The flux on the other three spheres also shows that there are two
positive maxima and negative minima near the midplane, which implies the
formation of a two-arm spiral structure in the circumplanetary disk.
The range of the elevation angle $\theta$ that corresponds to outward
flux shrinks with decreasing $\tilde{r}$, which indicates the change of
disk aspect ratio with $\tilde{r}$.

Although Figure \ref{Fig_flux_at_spheres} shows that the gas at high
elevation angle is always falling radially inward, it is difficult to
judge the direction of net mass flux in the disk (corresponding to low
$|\theta|$) from these plots.
In order to examine the net mass flux, next we calculate azimuthally
integrated mass flux at spheres with several radii as a function of
$\theta$ (Figure \ref{Fig_flux_vs_theta}(a)) :
\begin{equation}
\tilde{F}_r(\tilde{r},\theta)
= \int_0^{2\pi}
      (\tilde{\rho} \tilde{v}_r) \tilde{r}^2 \cos\theta d\phi.
\end{equation}
We find that the direction of the net radial flux near the midplane
($|\theta| \lesssim 20^\circ$ in the case of $\tilde{r}=1.0$), where the
gas moves both inward and outward through the sphere in the region
depending on the longitude, is actually outward, while it moves inward
for higher elevation angle ($|\theta| \gtrsim 20^\circ$ in the case of
$\tilde{r}=1.0$), which is obvious in Fig.~\ref{Fig_flux_at_spheres}.
This basic profile does not change with $\tilde{r}$, but where
$\tilde{r}\lesssim 0.1$, the profile becomes sharper, that is, the
region where gas moves outward is limited in narrower elevation angle,
which corresponds to the dependence of disk aspect ratio on radius.
This clear outward motion around the midplane for wide range of
$\tilde{r}$ shows another evidence of the unbound state of the gas where
$\tilde{r} \gtrsim 0.1$, which is already shown in the above
(Fig.~\ref{Fig_streamlines_outward}).
In the cases with $\tilde{r}=0.1$ and 0.03, there is a narrow band where
the mass flux is inward (negative) with peak at $\theta \sim \pm
25^\circ$ and $\pm 15^\circ$, respectively.
Figure \ref{Fig_flux_vs_theta}(b) shows azimuthally averaged radial
velocity:
\begin{equation}
\bar{\tilde{v}}_r(\tilde{r},\theta)
 \equiv
   \frac{\int_0^{2\pi} \tilde{\rho} (\tilde{r}, \theta, \phi)\,
                       \tilde{v}_r  (\tilde{r}, \theta, \phi) d\phi}
        {\int_0^{2\pi} \tilde{\rho} (\tilde{r}, \theta, \phi) d\phi}.
\end{equation}
This quantitatively shows that the velocity of the gas elements
accreting toward the disk nearly vertically at high elevation angle
($|\theta| \gtrsim 40^\circ$ in the case with $\tilde{r}=0.1$) is
nearly free-fall velocity ($\sim \sqrt{6\tilde{r}_{\rm
H}^3/\tilde{r}}$), which is faster than the sound speed ($=1$).
%
Near the midplane, radial velocity is very small, which corresponds to
the hydrostatic disk region described above
(Fig.~\ref{Fig_diskstructure}).  Between the two regions, there is a
transition region, which is under the shock standing at the disk
surface, and the gas in the region is not in dynamically equilibrium
state as a part of the disk.  The range of $\theta$ of the transition
region corresponds to the narrow band with inward mass flux described
above.  Thus, this is a kind of layered accretion and its mechanism will
be discussed in \S~\ref{sec_distribution}.

Figure \ref{Fig_vr-per-vKep} shows radial velocity in the midplane
normalized by local Keplerian velocity defined by
\begin{equation}
{\cal V}_R(\tilde{R})
 = \frac{\bar{\tilde{v}}_r(\tilde{r}=\tilde{R},\theta=0)}
        {\sqrt{3\tilde{r}_{\rm H}^3/\tilde{R}}}.
\label{vr_per_vKep}
\end{equation}
We find that radial velocity is very small but takes on positive values
for a wide range of radii ($0.005 \lesssim \tilde{R} \lesssim 0.2$).
Significant positive values at $\tilde{R} \gtrsim 0.2$ correspond to the
outward motion which is already shown in
Fig.~\ref{Fig_streamlines_outward}.
Non-negligible positive values of ${\cal V}_R$ at $\tilde{R}
\lesssim 0.005$ may reflect non-steady state small-scale structure in
the region, or artifact of averaging operation over discrete grids whose
size may not be sufficiently small as compared to the distance from the
origin.
Formation of the circumplanetary disk with nearly Keplerian motion is
also demonstrated in Fig.~\ref{Fig_AM_vs_r}, which shows azimuthally
averaged specific angular momentum at the midplane.  We clearly see that
the rotation velocity is nearly Keplerian for a wide range of radii
($\tilde{R} \lesssim 0.1$).
Keplerian rotation is realized when the ratio of thermal energy of the
gas to potential energy of the planet is much smaller than unity (or
pressure force is much weaker than gravity force).  Eq.~(\ref{h_p})
indicates that $\tilde{h}_{\rm p}/\tilde{R}$, which is square root of
the ratio, becomes smaller with decreasing $\tilde{R}$, which is
consistent with our result.

Figure \ref{Fig_rho_and_sigma_ave} shows the plots of azimuthally
averaged density at the midplane and surface density.  We can see that
the profiles at $\tilde{R} \gtrsim 0.02$ can be fitted roughly by
power-law functions as $\bar{\tilde{\rho}} \propto \tilde{R}^{-3}$
and $\bar{\tilde{\Sigma}} \propto \tilde{R}^{-3/2}$, respectively.  In
an equilibrium state, we have $\tilde{\rho} =
\tilde{\Sigma}/(\sqrt{2\pi} \tilde{h}_{\rm p})$.  The above two
functions imply that this relationship is approximately satisfied and
the hydrostatic equilibrium is realized.
Deviation from the power-law functions for the inner region may be due
to insufficient computation time as compared to the time required to
reach a steady state\footnote{Time required to reach steady state can be
estimated as $(\pi \tilde{R}^2 \bar{\tilde{\Sigma}} /
\tilde{\dot{M}})_{r=0.02} \sim 30$, which is longer than the duration
with full level calculation.  Here we used $\tilde{\dot{M}} =
\tilde{\dot{M}}_{\rm s}(\tilde{R})$ (see Fig.~\ref{Fig_M_s_M_Kep}).}, or
due to the effect of sink cells around the center.
Here we emphasize again that the distributions of mass and angular
momentum of accreting gas that we present in \S~\ref{sec_distribution}
are more important and useful quantities less affected by numerical
procedures, as we mentioned in \S~\ref{sec_introduction}.
Note that \citet{Machida09} introduced an idea of {\it centrifugal
barrier} to explain the peak of surface density.  However, we do not
find any physical reason for the existence of such a barrier because, as
we will show in \S \ref{sec_distribution}, distributions of accreting
mass and angular momentum onto the circumplanetary disk are well
described by power-law functions, which do not have typical lengths,
such as centrifugal radius.

Figures \ref{Fig_Bernoulli_wide} and \ref{Fig_Bernoulli_closeup} show
variation of physical quantities along two streamlines starting from
different heights that correspond to the passing and accreting regions.
Overall variation is shown in Fig.~\ref{Fig_Bernoulli_wide}, while
close-up view near the shock front is shown in
Fig.~\ref{Fig_Bernoulli_closeup}.  The key quantity in this figure is
the Bernoulli integral $\tilde{B}$ given by
\begin{equation}
\tilde{B}
 = \frac{1}{2}|\tilde{\BF{v}}|^2 + \log \tilde{\rho} + \tilde{\Phi},
\label{Bernoulli}
\end{equation}
where we included potential energy.  This quantity is conserved along
each streamline except shock surfaces.

\begin{figure}
\epsscale{.90}
\plotone{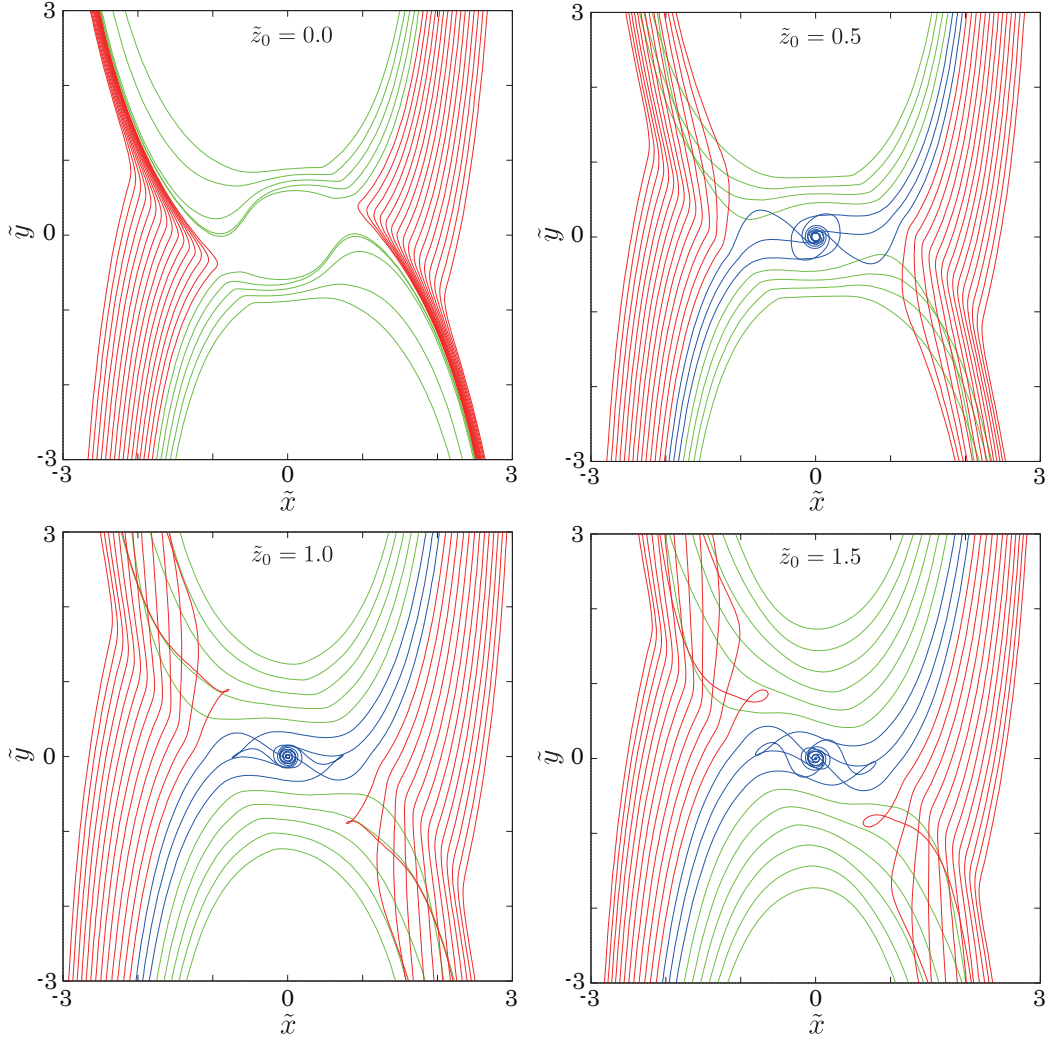}
%
\caption{Streamlines starting from four different heights ($\tilde{z}_0
= 0.0, 0.5, 1.0, 1.5$), with $\tilde{x}_0 = [\pm 2, \pm3]$ and
$\tilde{y}_0 = \pm \tilde{L}_y/2$.  Interval of the starting points is
0.05 in the $x$-direction.  Green, blue, and red curves show streamlines
in the horseshoe, accretion, and passing region, respectively.
\label{Fig_streamlines}}
\end{figure}

\begin{figure}
\epsscale{0.9}
\plotone{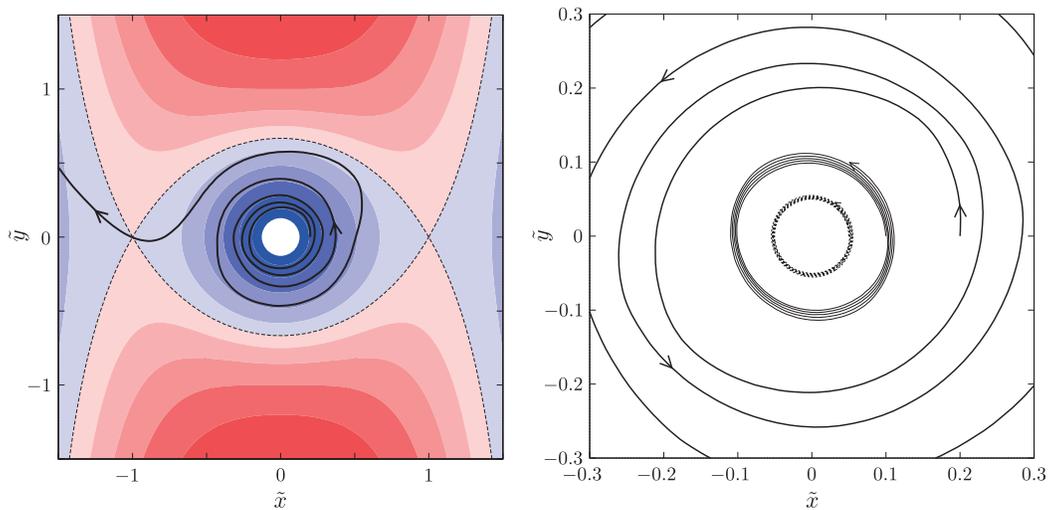}
%
\caption{Streamlines in the midplane of a circumplanetary disk.  Left
 panel shows a streamline starting from $\tilde{R}=0.2$ in the midplane
 $(\tilde{x}_0, \tilde{y}_0, \tilde{z}_0) = (0.2,0,0)$.  Dashed line
 shows contour line of $\tilde{\Phi}=0$, which passes the two Lagrangian
 points L$_1$ $(-1,0,0)$ and L$_2$ $(1,0,0)$.  Colors show potential
 $\tilde{\Phi}$; red regions are $\tilde{\Phi}>0$ and blue regions are
 $\tilde{\Phi}<0$.  Right panel shows a closer view of three streamlines
 starting from $\tilde{R}=0.2$ (same as the left panel), 0.1, and 0.05,
 respectively.
\label{Fig_streamlines_outward}}
\end{figure}

First we examine a streamline in the midplane plane
(Figs.~\ref{Fig_Bernoulli_wide}(a) and \ref{Fig_Bernoulli_closeup}(a)).
The starting point of the streamline is $(\tilde{x}_0, \tilde{y}_0,
\tilde{z}_0) = (2.37, \tilde{L}_y/2, 0)$, which is in the passing region
(red region in Figs.~\ref{Fig_streamlines} and \ref{Fig_fate});
$\tilde{x}$ keeps positive values not far from 2, and $\tilde{y}$
monotonically decreases with increasing $\tilde{d}$, where $\tilde{d}$
is the distance from the starting points of streamlines.
With approaching to the planet (corresponding to $\tilde{d} \lesssim
11.6$), the gas element climbs the tidal potential slope decreasing its
kinetic energy, and density slightly decreases.  Although these
quantities change with $\tilde{d}$, Bernoulli integral $\tilde{B}$,
which is the sum of the three quantities, is constant until $\tilde{d}
\sim 11.6$, which shows there are no shocks along the path until this
point.
However, there is a shock surface at $\tilde{d} \sim 11.6$ where kinetic
energy decreases and density increases discontinuously, and $\tilde{B}$
decreases as a result of shock dissipation.
This shock surface stands from the Hill sphere toward both inside and
outside of the planet orbit, forming spiral density waves around the
central star.
After passage of the shock, $\tilde{B}$ remains nearly constant,
although other quantities change with increasing $\tilde{d}$.  This
means that the gas element on the streamline underwent only one shock
with modest strength.

Next we examine an off-midplane streamline
(Figs.~\ref{Fig_Bernoulli_wide}(b) and \ref{Fig_Bernoulli_closeup}(b)).
The starting point is $(\tilde{x}_0, \tilde{y}_0, \tilde{z}_0) = (2.37,
\tilde{L}_y/2, 1)$, which is just above the previous starting point with
the same horizontal position, and this corresponds to the accreting
region (blue region in Figs.~\ref{Fig_streamlines} and \ref{Fig_fate}).
The quantities change in a similar way to the case in the midplane
before $\tilde{d} \sim 12.0$, where the gas element hits the shock
surface forming around the Hill sphere.  We have non-zero $\tilde{z}$
values in this case but it is kept almost constant until it encounters
the shock, which shows that the flow before the shock is laminar.
When the gas element passes through the shock surface at $\tilde{d} \sim
12.0$, basic behaviour is the same as the $\tilde{z}_{\rm 0}=0$ case,
but the gas element changes the direction slightly upward (positive
$\tilde{z}$ direction) at the shock (see the line showing the change in
$\tilde{z}$ in Fig.~\ref{Fig_Bernoulli_closeup}(b)).  This is because
the shock surface forms a bow shock and is curved upward, thus the gas
element of off-midplane, which passed an oblique shock, gains upward
momentum.
After that, the gas element falls steeply toward the mid-plane and hits
the surface of the circumplanetary disk, forming another shock at
$\tilde{d} \sim 13.9$.
The point of fall is at $\tilde{r} \sim 0.1$, which is sufficiently
close to the planet to make use of its gravity for acceleration.  Thus
Mach number of the gas when it hits the disk surface is much larger than
unity, and strong shock dissipation is caused.  Owing to this strong
dissipation, the gas becomes captured within the Hill sphere and merged
as a part of disk.
Note that the gradual decrease of $\tilde{B}$ before $\tilde{d} \sim
13.9$ arises presumably from collisions between falling gas elements on
the way to the surface, especially between two gas elements originated
from $\tilde{x}>0$ and $\tilde{x}<0$ in the protoplanetary disk (see also
Fig.~\ref{Fig_flux_disk-surface} and the corresponding description in
the text).
After the passage of the shock forming above the circumplanetary disk
($\tilde{d} \gtrsim 14$), $\tilde{r}$ becomes almost constant and
azimuth angle changes with a constant rate, which means that the gas
is now rotating around the planet, i.e., the gas element has become a
part of the circumplanetary disk.
Note that specific angular momentum is not constant along the
streamlines at all, although it is sometimes assumed to be a conserved
quantity.

%
Now we consider why the gas in the midplane cannot accrete to the
circumplanetary disk.
As described above, effective shock dissipation is required for gas
elements to become accreted onto the circumplanetary disk, and the
planet gravity is essential for the effective acceleration.
When gas element consumes potential energy without significant increase
of gas density, kinetic energy of the gas increases effectively (see
Eq.~(\ref{Bernoulli})) and the gas element can have a high Mach number,
which leads to strong shock dissipation.
However, in order for the gas in the midplane to consume planet
gravitational energy, the gas element has to pass through the high
density region of the circumplanetary disk.  The increase of density
absorbs potential energy and prevents gas from effective acceleration.
Gas near the midplane is thus difficult to form strong shocks and
difficult to dissipate energy effectively, which prevents it from being
captured by the planet gravity.  Therefore, the gas in the midplane is
difficult to be accreted into the circumplanetary disk.

One may think that passing many weak shocks during rotating around the
planet in the circumplanetary disk would contribute to inward migration
of the rotating gas.
However, the decrease in $\tilde{B}$ is a third-order small quantity
with respect to $({\cal M}-1)$ (where ${\cal M}$ is Mach number), so
such weak shocks are not effective for energy dissipation and inward
migration.
Note also that acceleration by the tidal force in the $z$-direction
alone is not sufficiently strong to form strong shocks.  This is because
the gas is in hydrostatic equilibrium supported by thermal pressure, so
that the resultant accelerated velocity should not be much larger than the
sound speed, being  able to produce only weak shocks.
We therefore conclude that the strong shock produced by direct infall
onto the surface of the inner part of the circumplanetary disk and
associated energy dissipation is essential for the accretion of the gas
from protoplanetary disk onto the circumplanetary disk.

\subsection{Distribution of Accreting Mass and Angular Momentum onto the
  Disk}\label{sec_distribution}

Next, we examine accretion rate of mass and angular momentum as a
function of distance from the planet.
First we define the surface at which physical quantities immediately
prior to the accretion onto the surface of a circumplanetary disk is
measured.  The surface is axisymmetric about the $z$-axis and the height
of the surface from the midplane is denoted by $\tilde{z}_{\rm
s}(\tilde{R})$.  We define the height of the surface by a linear
function of the local scale height of the circumplanetary disk as
\begin{equation}
\tilde{z}_{\rm s} (\tilde{R})
 = f_h \tilde{h}_{\rm p} + \tilde{z}_{\rm s0},
\label{z_s}
\end{equation}
where $f_h$ and $\tilde{z}_{\rm s0}$ are constants.  We set $f_h = 6$
and $\tilde{z}_{\rm s0} = 0.004$ so that the surface $\tilde{z}_{\rm s}$
is above but not far from the shock surface of the circumplanetary disk
in the region with $\tilde{R}<0.2$.

Next we define mass flux onto the circumplanetary disk by
\begin{equation}
\tilde{f}_{\rm s} (\tilde{R},\phi)
 \equiv - \frac{\tilde{\rho}_{\rm s} \tilde{\BF{v}}_{\rm s}
                \cdot \BF{n}_{\rm s}}
               {\cos \theta_{\rm s}},
\label{f_s}
\end{equation}
where $\tilde{\rho}_{\rm s}$ and $\tilde{\BF{v}}_{\rm s}$ are the
density and velocity vector at the surface defined by Eq.~(\ref{z_s}),
$\BF{n}_{\rm s}$ is an upward unit vector normal to the surface,
$\theta_{\rm s} = d\tilde{z}_{\rm s}/d\tilde{R}$, and the negative sign
is added so that $\tilde{f}_{\rm s}$ becomes positive when the gas
passes the surface downward.
Note that $\cos \theta_{\rm s}$ appears in the denominator in
Eq.~(\ref{f_s}) because $\tilde{f}_{\rm s}$ is defined as a flux at the
surface per unit area of the $x$--$y$ plane, not a flux per unit area of
the surface.

Figure \ref{Fig_flux_disk-surface} shows several quantities on the
surface $\tilde{z} = \tilde{z}_{\rm s}$ projected on the
$\tilde{x}$--$\tilde{y}$ plane.
We find that the density of the accreting gas at the surface
$(\tilde{\rho}_{\rm s})$ has peaks at $(\tilde{x},\tilde{y}) \simeq (\pm
0.1, \mp 0.08)$, while the mass flux $\tilde{f}_{\rm s}$ has peaks at
$(\tilde{x},\tilde{y}) \simeq (\pm 0.04, \mp 0.04)$, which are located
closer to the planet than the density maxima.  This is because accreting
velocity, which is almost free-fall velocity, is faster at the inner
region.
These two peaks of the two-arm structure correspond to the places where
gas elements coming from interior ($\tilde{x}<0$) and exterior
($\tilde{x}>0$) to the planet orbit collide with each other.
As for the $z$-component of specific angular momentum around the planet
measured in the inertial frame $\tilde{j}_{z,\rm s} \equiv
(\tilde{\BF{r}} \times \tilde{\BF{v}})_{{\rm at}\,
\tilde{z}=\tilde{z}_{\rm s} } \cdot \BF{e}_z + \tilde{R}^2$ (where the
$\tilde{R}^2$ term arises from the rotation of the coordinate system),
its value is positive (i.e., prograde) in the whole region.  The
distribution is roughly axisymmetric with some elongation in the
$x$-direction within $\tilde{R} \lesssim 0.1$, and the value increases
with increasing radius.  As for the angular momentum flux, the above
axisymmetric property breaks and has peaks at $(\tilde{x},\tilde{y})
\simeq (\pm 0.12, \mp 0.08)$, which simply corresponds to the two-arm
structure observed in density or mass flux.
Note that the negative values observed in $\tilde{f}_{\rm s}$ in the
outer region ($\tilde{R} \gtrsim 0.1$) are due to the fact that the gas
passes through the $\tilde{z} = \tilde{z}_{\rm s}$ surface from beneath.
This is because, in this region $(\tilde{R} \gtrsim 0.1)$, the ratio of
the scale height to the radial distance of the circum-planetary disk is
not much smaller than unity anymore and $\tilde{z}_{\rm s}$ is
comparable to $\tilde{R}$ accordingly.  This leads to negative
$\tilde{f}_{\rm s}$ for a gas element whose horizontal velocity is
comparable or larger than vertical velocity even when the vertical
velocity is negative (downward).
The surface of the disk in this region is hard to define, thus the way
to measure mass flux described above may not be appropriate in this
region.

For a better understanding of the distribution of accreting gas, we next
show azimuthally-averaged quantities at the surface defined by
Eq.~(\ref{z_s}).
%
First we examine mass accretion rate.  Figure \ref{Fig_f_s_f_Kep} shows
the plots of azimuthally-averaged mass flux onto the disk, defined by
\begin{equation}
\bar{\tilde{f}}_{\rm s}(\tilde{R})
 \equiv \frac{1}{2\pi}
        \int_0^{2\pi}
            \tilde{f}_{\rm s}(\tilde{R}, \phi)
        d\phi.
\label{f_s_bar}
\end{equation}
Figure \ref{Fig_M_s_M_Kep} shows the plots of the cumulative mass
accretion rate, i.e., the mass accreted per unit time through the part
of the surface $\tilde{z}_{\rm s}$ where horizontal distance
from the planet is smaller than $\tilde{R}$, i.e.,
\begin{equation}
\dot{\tilde{M}}_{\rm s}(\tilde{R})
 \equiv \int_0^{\tilde{R}}
        \int_0^{2\pi}
            \tilde{f}_{\rm s}(\tilde{R}, \phi)
            \tilde{R}' d\phi\, d\tilde{R}'.
\label{M_s_dot}
\end{equation}
We can see from Fig.~\ref{Fig_f_s_f_Kep} that $\bar{\tilde{f}}_{\rm
s}(\tilde{R})$ is almost constant from $\tilde{R} \sim 0.1$ all the way
to the very center ($\tilde{R}\sim 0.001$).  Cumulative mass accretion
rate $\dot{\tilde{M}}_{\rm s}$ is thus roughly proportional to
$\tilde{R}^2$, which is a power-law function and thus does not have any
typical length except for the outer end of the downward accretion on the
disk surface (at $\tilde{R} \sim 0.1$ in this case).
This distribution shows that the main contribution of mass accretion
comes from the outer region ($\tilde{R} \sim 0.1$) and there is no
specific radius where gas accretion is concentrated somewhere in
between.
Note that dotted lines in Figures \ref{Fig_f_s_f_Kep} and
\ref{Fig_M_s_M_Kep} show quantities calculated by Eqs.~(\ref{f_s_bar})
and (\ref{M_s_dot}), respectively, integration being performed over only
regions with $\tilde{f}_{\rm s}>0$.  Deviation from each solid line can
be only seen at $\tilde{R} \gtrsim 0.15$, where gas can pass through the
$\tilde{z} = \tilde{z}_{\rm s}$ surface even upward depending on azimuth
angle (see Fig.~\ref{Fig_flux_disk-surface}).
This is the reason why $\dot{\tilde{M}}_{\rm s}$ (solid line) decreases
with increasing radius at $\tilde{R} \gtrsim 0.2$ although it is a
cumulative quantity.
The treatment of our analysis in this region is a difficult problem and
will be discussed later (\S \ref{sec_disk_size}).

The distribution of $\bar{\tilde{f}}_{\rm s}$ (or $\dot{\tilde{M}}_{\rm
s}$) provides us with useful information but it is not sufficient for an
understanding of the mass distribution in a circumplanetary disk.
In general, the radial distance from the planet of the location where
gas is first accreted on the disk is different from its final orbit
after it settles in as a part of the Keplerian rotating disk.
What connects the two is angular momentum of the gas.
Figure \ref{Fig_j_zs} shows azimuthally-averaged specific angular
momentum $\bar{\tilde{j}}_{z,\rm s}$ at the $\tilde{z} =
\tilde{z}_{\rm s}$ surface given by
\begin{equation}
\bar{\tilde{j}}_{z,\rm s} (\tilde{R})
= \frac{\displaystyle
        \int_0^{2\pi} \tilde{j}_{z,\rm s}
                      \tilde{\rho}_{\rm s} d\phi}
       {\displaystyle
        \int_0^{2\pi} \tilde{\rho}_{\rm s} d\phi}.
\end{equation}
We can see that $\bar{\tilde{j}}_{z,\rm s}$ is always smaller than the
specific angular momentum for Keplerian rotation $\tilde{j}_{\rm Kep} =
\sqrt{3\tilde{r}_{\rm H}^3\tilde{R}}$.
This means that the gas does not have enough angular momentum for
Keplerian rotation at the radius, which would lead inward migration of
the gas after the fall on the disk surface, while the gas near the
midplane moves outward, as mentioned above (Fig.~\ref{Fig_vr-per-vKep}).
%
This explains the inward stream of the layer just under the shock
of the disk surface (Fig.~\ref{Fig_flux_vs_theta}).
In addition, $\bar{\tilde{j}}_{z,\rm s}$ is nearly proportional to
$\tilde{R}$, which is steeper than $\tilde{j}_{\rm Kep} \propto
\tilde{R}^{1/2}$, and thus the ratio of $\bar{\tilde{j}}_{z,\rm s}$ to
$\tilde{j}_{\rm Kep}$ decreases with decreasing $\tilde{R}$.
This suggests that the gas accreted in the outer region of the disk
($\tilde{R} \sim 0.1$) tends to keep the position when it rotates as a
part of the circumplanetary disk, whereas the gas accreted in the inner
region tends to move further inward in order to achieve balance between
centrifugal and gravitational forces.

We now have the mass flux and angular momentum in the accretion flow,
which allows us to estimate the {\it effective} distribution of
accreting gas elements, assuming their redistribution to radial
distances where their specific angular momentum matches that of the
local Keplerian rotation.
Let $\tilde{R}_{\rm Kep}$ denote the radius where the gas with specific
angular momentum $\tilde{j}_{z,\rm s}$ is rotating with the Keplerian
velocity, i.e.,
\begin{equation}
\tilde{R}_{\rm Kep} (\tilde{R},\phi)
 = \frac{\tilde{j}_{z, \rm s}^2}{3\tilde{r}_{\rm H}^3}.
\end{equation}
Then the effective distribution of azimuthally-averaged mass flux after
redistribution of the gas with angular momentum conservation can be
written as
\begin{equation}
\bar{\tilde{f}}_{\rm Kep}(\tilde{R}) d\tilde{R}
= \frac{1}{2\pi \tilde{R}}
  \int_0^\infty d\tilde{R}'
  \int_0^{2\pi} \tilde{R}' d\phi \,
  \tilde{f}_{\rm s}(\tilde{R}',\phi)
  \delta ({\tilde{R}',\phi}),
\label{f_Kep_bar}
\end{equation}
where
\begin{eqnarray}
\delta(\tilde{R}',\phi) =
\begin{cases}
 1 & \mbox{if $\tilde{R}-d\tilde{R}/2
               < \tilde{R}_{\rm Kep}(\tilde{R}',\phi)
               < \tilde{R}+d\tilde{R}/2$
           and $\tilde{f}_{\rm s}>0$}, \\
 0 & \mbox{otherwise}.
\end{cases}
\end{eqnarray}
The corresponding cumulative mass accretion rate is
\begin{equation}
\dot{\tilde{M}}_{\rm Kep}(\tilde{R})
 = \int_0^{\tilde{R}}
       2\pi \tilde{R}' \bar{\tilde{f}}_{\rm Kep}(\tilde{R}')
       d\tilde{R}'.
\label{M_Kep_dot}
\end{equation}
The plots of $\bar{\tilde{f}}_{\rm Kep}$ and $\dot{\tilde{M}}_{\rm Kep}$
are shown in Figures \ref{Fig_f_s_f_Kep} and \ref{Fig_M_s_M_Kep}.
We can see that $\bar{\tilde{f}}_{\rm Kep}$ is roughly proportional to
$\tilde{R}^{-1}$ and $\dot{\tilde{M}}_{\rm Kep}$ is to $\tilde{R}$.
By comparison with $\bar{\tilde{f}}_{\rm s}$, we find that
$\bar{\tilde{f}}_{\rm Kep}$ is more center-concentrated distribution,
but still outer region is a dominant source of gas accretion in the
sense that $\partial \ln \tilde{f}_{\rm Kep}/\partial \ln \tilde{R} >
-2$.
Since angular momentum of a gas element after the shock of the disk
surface is not necessarily conserved until the gas reaches the radius
where it rotates in the Keplerian velocity, the distributions of
$\bar{\tilde{f}}_{\rm Kep}$ and $\dot{\tilde{M}}_{\rm Kep}$ should be
regarded as the case of the most center-concentrated limit.



\section{Discussion} \label{sec_discussion}

\subsection{Size of Circumplanetary Disks}
\label{sec_disk_size}
The size of a circumplanetary disk is closely related to the location of
satellite formation.  However, it is not easy to define the outer edge
of the disk, and there have been several attempts.
One natural way would be to define the disk edge based on its density
distribution.  But the density is monotonically and smoothly decreasing
with increasing radius toward the Hill sphere, so it is difficult to
define the edge from density distribution.
\citet{Ayliffe09b}, who performed three-dimensional hydrodynamic
simulations, suggested a criterion for the disk edge based on angular
momentum distribution.  Their simulations show that specific angular
momentum of the disk gas has a peak at a certain radial location,
whereas specific angular momentum of a Keplerian disk increases
monotonically.  Their Figure 2 suggests that the turnover point is about
1/3 of the Hill radius, and they defined the disk edge by the radial
location of this peak.
\citet{Martin11} examined periodic orbits of a particle around a planet
under the influence of the gravitational force from a central star and
the planet, and found that, as the size of the orbit is increased, the
orbits start crossing with each other at $r \sim 0.4r_{\rm H}$.  They
inferred that this corresponds to the location of the disk's outer edge
where tidal torque of central star's gravity becomes strong, and called
it tidal truncation radius ($r_{\rm trunc}$).
The above value of $r_{\rm trunc}$ is in agreement with the point of
turnover of specific angular momentum found in their two-dimensional SPH
simulation, as well as in the result of \citet{Ayliffe09b}.
Our Figure \ref{Fig_AM_vs_r} shows that the turnover point is $\tilde{R}
\sim 0.3 \tilde{r}_{\rm H}$, which is also roughly in agreement with
$r_{\rm trunc}$.\footnote{In this comparison, we assume that specific
angular momentum in these works were not measured in inertial frame, but
in the rotating frame.}

We here consider an alternative criterion by examining ${\cal V}_R$
defined by Eq.~(\ref{vr_per_vKep}) to define the position of the outer
edge.
We observed that azimuthally-averaged radial velocity in the midplane is
positive (outward) in almost all the region, although the value is very
small for $\tilde{R} \lesssim 0.2$ (Fig.~\ref{Fig_vr-per-vKep}), and
the outward velocity significantly increases at $\tilde{R} \gtrsim 0.2$
(see also Fig.~\ref{Fig_streamlines_outward}).
Since gas at $\tilde{R} \gtrsim 0.2$ moves outward and escapes from the
Hill sphere quickly, the region $\tilde{R} \gtrsim 0.2$ can be regarded
as outside of the circumplanetary disk.
We thus define the disk edge as the radial location where ${\cal V}_R$
starts increasing significantly and has non-negligible positive value;
$\tilde{R} \sim 0.2$ in our case shown in Fig.~\ref{Fig_vr-per-vKep}.
%
Note that a two-dimensional SPH simulation shows that negative torque is
exerted on the outer disk \citep{Martin11}, which gives rise to inward
flow and is contrary to our results.  This might be due to the fact that
two-dimensional simulations, which create clearer spiral structure than
three-dimensional ones, tend to enhance torque density.
In particular, the disk thickness in the outer region ($\tilde{R}
\gtrsim 0.2$), at which outflow is observed, is very thick (thickness is
comparable to radius).  Thus, it would be unlikely that our
three-dimensional calculation is significantly affected by the negative
torque suggested by \citet{Martin11}.
%
%
%
The distribution of accreting angular momentum also seems to indicate a
similar radius for the edge.  As we see in Fig.~\ref{Fig_j_zs},
$\bar{\tilde{j}}_{z,\rm s}$ and $\tilde{j}_{\rm Kep}$ have different
dependence on $\tilde{R}$.  If we fit each profile as a single
power-law function, they cross each other at $\tilde{R}\sim 0.3$ in the
present case.
This roughly agrees with the location of the disk edge defined above
based on the significant increase of outward velocity.
The gas exterior to this radius has too large angular momentum to
achieve Keplerian rotation, thus moves radially outward.
%

%
Once angular momentum of accreting gas is obtained, one can calculate
mean specific angular momentum of the accretion flow, which is sometimes
used in calculating the the so-called centrifugal radius $\tilde{r}_{\rm
c} \equiv \tilde{\ell}^2/(3\tilde{r}_{\rm H}^3)$ to infer the disk size
\citep[e.g.,][]{Mosqueira03a,Ward10}.
Mean specific angular momentum of the accretion flow within a radius
$\tilde{R}$ is given by
\begin{equation}
\tilde{\ell}(\tilde{R})
 = \frac{\dot{\tilde{J}}_{\rm s}(\tilde{R})}
        {\dot{\tilde{M}}_{\rm s}(\tilde{R})},
\end{equation}
where
\begin{equation}
\dot{\tilde{J}}_{\rm s}(\tilde{R})
 = \int_0^{\tilde{R}}
       d\tilde{R}'
   \int_0^{2\pi}
       \tilde{R}'d\phi
       \tilde{f}_{\rm s}(\tilde{R}',\phi) \tilde{j}_{\rm s}(\tilde{R}',\phi).
\end{equation}
We can see from Figure \ref{Fig_ell} that $\tilde{\ell}(\tilde{R})$
increases nearly linearly with radius and levels around $\tilde{R} \sim
0.2$, where $\tilde{\ell} \sim 0.2$.  This corresponds to
$\tilde{r}_{\rm c} \sim 0.013$, which might give us a rough estimation
of the disk size.
However, it is practically difficult to determine the upper bound of the
integral range with respect to $\tilde{R}$, which affects the value of
$\tilde{\ell}$, because the circumplanetary disks is smoothly connected
to the proto-planetary disks.
Also, unlike the case of a particle, angular momentum of a gas element
is not a conserved quantity along a flow in principle, because gas is
continuum medium which can transfer angular momentum through waves.
This also makes it difficult to determine $\tilde{\ell}$ precisely, and
thus $\tilde{r}_{\rm c}$ as well.
%



\subsection{Picture of Gas Accretion Flow onto Circumplanetary Disks}
\label{sec_picture}
Figure \ref{Fig_CP-disk} shows a schematic picture of gas accretion flow
onto a circumplanetary disk based on the results obtained in the present
work.  Gas accretion occurs mostly downward from high altitude with high
incident angle (see Fig.~\ref{Fig_diskstructure}).  The accreting gas is
accelerated by the planet gravity to have almost free-fall velocity of
the planet (Fig.~\ref{Fig_flux_vs_theta}b).  The value of angular
momentum of the accreting gas normalized by the local Keplerian angular
momentum is lower in the inner region of the disk and higher in the
outer region (Fig.~\ref{Fig_j_zs}).  The falling gas reaches the shock
surface formed on the top of the circumplanetary disk.

Gas near the midplane (especially where $\tilde{R} \lesssim 0.1$) is
almost in Keplerian rotation and hydrostatic equilibrium in the
$z$-direction.  In this region, radial velocity is very small and it is
difficult for the gas to accrete inward through the disk midplane, as
mentioned above.
On the other hand, the gas at $\tilde{R} \gtrsim 0.2$ in the midplane
shows significant outflow, which eventually escapes from the Hill sphere
(see Fig.~\ref{Fig_streamlines_outward}).
Thus inflow from high altitude and outflow near the midplane co-exist,
and they do not interfere with each other.  This means that there is a
circulation across the Hill sphere and fresh (protoplanetary) gas is
always supplied in the $\tilde{R} \gtrsim 0.2$ region.
In addition, the two Lagrangian points L$_1$ and L$_2$, which are often
thought to be the most likely points of inflow, are actually the points
of outflow, even in the gas accretion stage
(Fig.~\ref{Fig_flux_at_spheres}).
%
This seems consistent with \citet{Klahr06}, who also performed
three-dimensional hydrodynamic simulation with no explicit physical
viscosity, but with lower resolution and with radiation.  They showed
that accretion mainly occurs via the poles of the planet and no inflow
along the equatorial plane, which is quite similar to ours.  But they
explained that the circulation in their simulation is driven by
accretion heating, which we do not consider.  Thus it is not clear if
the driving mechanisms of the circulation are the same.
%
Note that \citet{Ayliffe09b} shows that the vast majority of the mass
flows into the Hill sphere near the equator, which seems to be
inconsistent with our result.  There are several differences in setting
between theirs and ours, so it is not easy to judge which factor causes
the difference.  Disk thickness would be one of the reasons for that,
even though our disk thickness is not very different from theirs.
Another possibility is gap formation.
When a deep gap is formed, the contribution of the vertically accreting
gas would become less significant, which reduces the difference.
However, here we think that viscosity is the main reason for the
difference.  When there is explicit viscosity, the circumplanetary disk
gas should transfer its angular momentum outward and most of gas would
move inward accordingly, but this is not necessarily true in the
inviscid limit.  Actually, inviscid simulations by \citet{Klahr06}
showed similar results with ours.  Also, inviscid limit might not be bad
because circumplanetary disks are likely to be MRI inactive in most
cases \citep{Fujii11}.

Since dust particles tend to settle down toward the midplane, gas
accretion flow from high altitude is likely dust-poor gas, which
diminishes dust-to-gas ratio in the circumplanetary disk.
On the other hand, since gas in the outflow region ($\tilde{R} \gtrsim
0.2$) rotates significantly slower than Keplerian velocity around the
planet, dust particles would migrate inward quickly.  Thus the outflow
in the midplane is also likely dust-poor gas, which would be the source
of solid material in the circumplanetary disk and would enhance
dust-to-gas ratio in the disk.
Dust-to-gas ratio in the circumplanetary disk is one of the most
important factors for satellite formation processes \citep{CW06}, thus
these two filtering effects would become important for satellite
formation.
In addition, studies on size evolution of solid material in
protoplanetary disks, such as \citet{Kobayashi10}, would also be
important for the filtering effects.
Further studies are needed on this issue.


As we see in Figure \ref{Fig_flux_vs_theta}, there is the layer just
under the shock at the disk surface where gas moves inward.  This {\it
layered accretion} is explained by the fact that angular momentum of the
accreting gas onto the disk surface is lower than that of the Keplerian
rotation at the radius (Fig.~\ref{Fig_j_zs}).
The inward stream in the layer might play an important role for the net
mass accretion toward the planet in circumplanetary disks.
This accretion picture should be explored more quantitatively by further
high-resolution simulations, using a code with lower artificial
viscosity such as the one with polar coordinates for numerical grids.

\subsection{Effects of Gap Formation}
One important issue to be addressed is the effect of gap, which is a
lower density annulus region near the planet orbit in the protoplanetary
disk.
We observed the gap as a slightly lower density band formed around
$\tilde{x}=0$ in our simulations.
However, in the last stage of giant planet formation, a giant planet
would become massive enough to create a deep gap, which would be able to
truncate its growth.
Since the deep gap is associated with steep density gradient at the
edge, gap formation may affect accretion flow and circumplanetary disk formation.
However, at the edge of such a deep gap, the gas density changes over a
radial distance comparable to the disk scale height, while the width of
the accretion band onto the circumplanetary disk is much narrower than
the scale height (Fig.~\ref{Fig_fate}).
Therefore, we think that the gap formation would not affect the
qualitative feature of the accretion flow.
Another possible effect of gap formation is the disk size.  As described
in \S \ref{sec_disk_size}, several mechanisms are proposed to explain
the disk size.  If the disk size is determined by the tidal effect as
suggested by \citet{Martin11}, it should not be affected by gap
formation, except when the Bondi radius is smaller than the Hill radius.
On the other hand, if the disk size is determined by the radially
outward velocity of the gas as described in \S \ref{sec_disk_size},
lower density around the Hill sphere and thus stronger radial pressure
gradient near the disk edge may enhance the outflow from the
circumplanetary disk, and the disk size may become smaller.
In any case, effects of gap formation should be examined in future works
to check the validity of the results shown in this paper.

%

\section{Summary} \label{sec_summary}

In order to understand the structure of circumplanetary disks, we
performed high-resolution hydrodynamic simulation and analyzed
gas accretion flow in detail.
We confirmed that gas accretion onto circumplanetary disks occurs in a
manner that the gas is accreted from high altitude toward the disk
surface downward with large incident angle, which was suggested by
previous studies \citep[e.g.,][]{DAngelo03,Bate03}, whereas
\citet{Ayliffe09b} showed that, in terms of mass flux, the downward
accretion flow is not significant, which is inconsistent with our
results.
We found that the gas that has passed through the shock surface moves
inward because its specific angular momentum is smaller than that of
Keplerian rotation, whereas gas accretion through the midplane does not
occur.
While the net gas accretion across the Hill sphere is inward, outflow
through near the two Lagrangian points is observed, although the regions
around the two Lagrangian points have been, from the point of view of
the potential energy, thought as a main accretion channel from
protoplanetary disks.
This outflow was not observed in previous hydrodynamic simulations for
Jupiter-sized planets \citep[e.g.,][]{Bate03,Ayliffe09b}.
Outward radial velocity of the gas near the midplane significantly
increases at a radial distance of about 0.2 times the Hill radius from
the planet, and the gas can escape from the Hill sphere within a short
period of time.

We also obtained the distribution of mass and angular momentum of
accreting gas onto the surface of circumplanetary disks.  We found that
the accretion rates of mass and angular momentum can be well described
by power-law functions.
This distribution would be useful in the study of satellite formation,
for example, a radially one-dimensional viscous-evolution model for
circumplanetary disks, such as \citet{Ward10} and \citet{Martin11}.
Recent development of viscous modeling for MRI turbulence in
protoplanetary and circumplanetary disks \citep{Okuzumi11, Fujii11}
would also contribute to construct more realistic models for the
circumplanetary disks.
However, in order to understand satellite formation processes, long term
evolution of circumplanetary disks, which is determined by the evolution
of accretion rate from protoplanetary disks to circumplanetary disks, is
necessary and thus global evolution of protoplanetary disks with
embedded giant planets until the complete dissipation of protoplanetary
disks needs to be examined.
Together with such global models \citep[e.g.,][]{Tanigawa07,Ward10}, we
will be able to obtain long-term evolution of circumplanetary disks,
which would provide better understanding of satellite formation
processes.

\begin{figure}
\epsscale{0.35}
\plotone{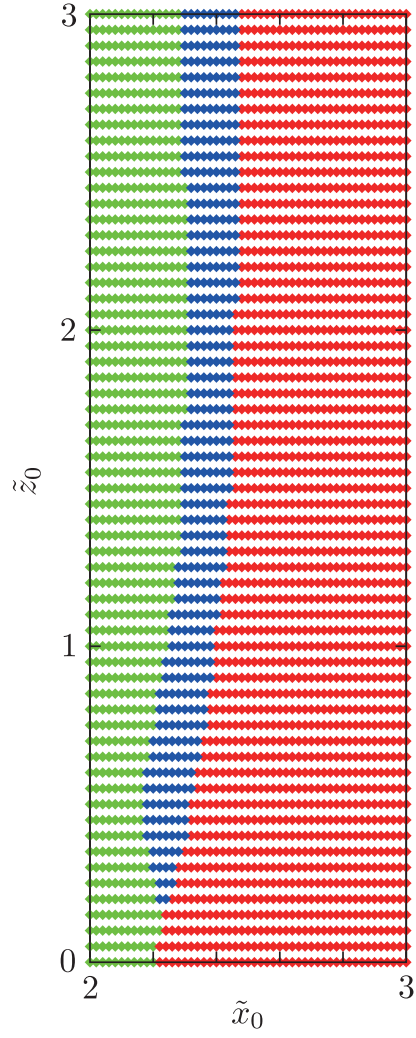}
%
\caption{Classification of starting points of streamlines on the
$x$--$z$ plane at $\tilde{y} = \tilde{L}_y/2$.  Streamlines starting
from the red region reach the boundary $\tilde{y}=-\tilde{L}_y/2$ with
$\tilde{x}>0$ (passing region), those in green region make U-turn and
reach the boundary $\tilde{y}=\tilde{L}_y/2$ with $\tilde{x}<0$
(horseshoe region), and those in blue region become trapped in
$\tilde{r}<0.2$ (accretion region).
\label{Fig_fate}}
\end{figure}

\begin{figure}
\epsscale{0.5}
\plotone{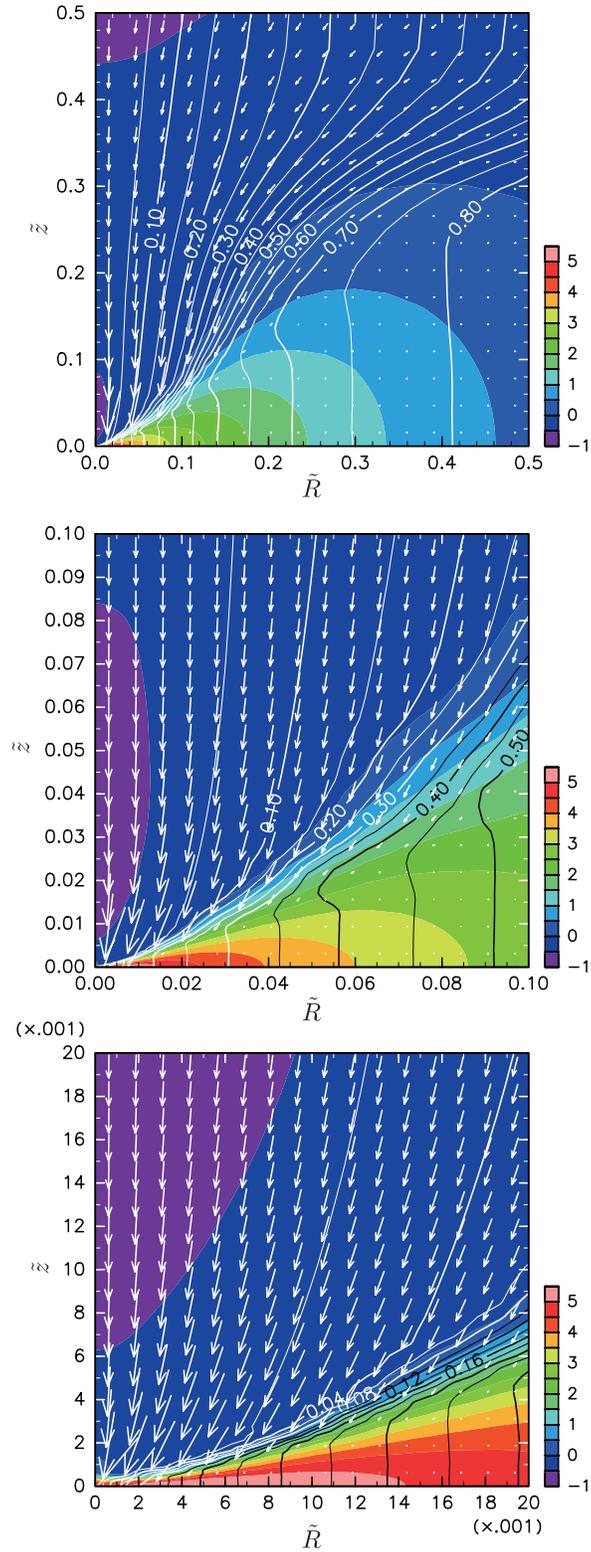}
%
\caption{Circumplanetary disk structure in the $R$--$z$ plane with three
 different spatial scales.  Log density is shown with colors, specific
 angular momentum is shown with contour lines, and gas velocity is
 expressed with arrows.  All quantities are averaged in the azimuthal
 direction.
\label{Fig_diskstructure}}
\end{figure}

\begin{figure}
\epsscale{1.0}
\plotone{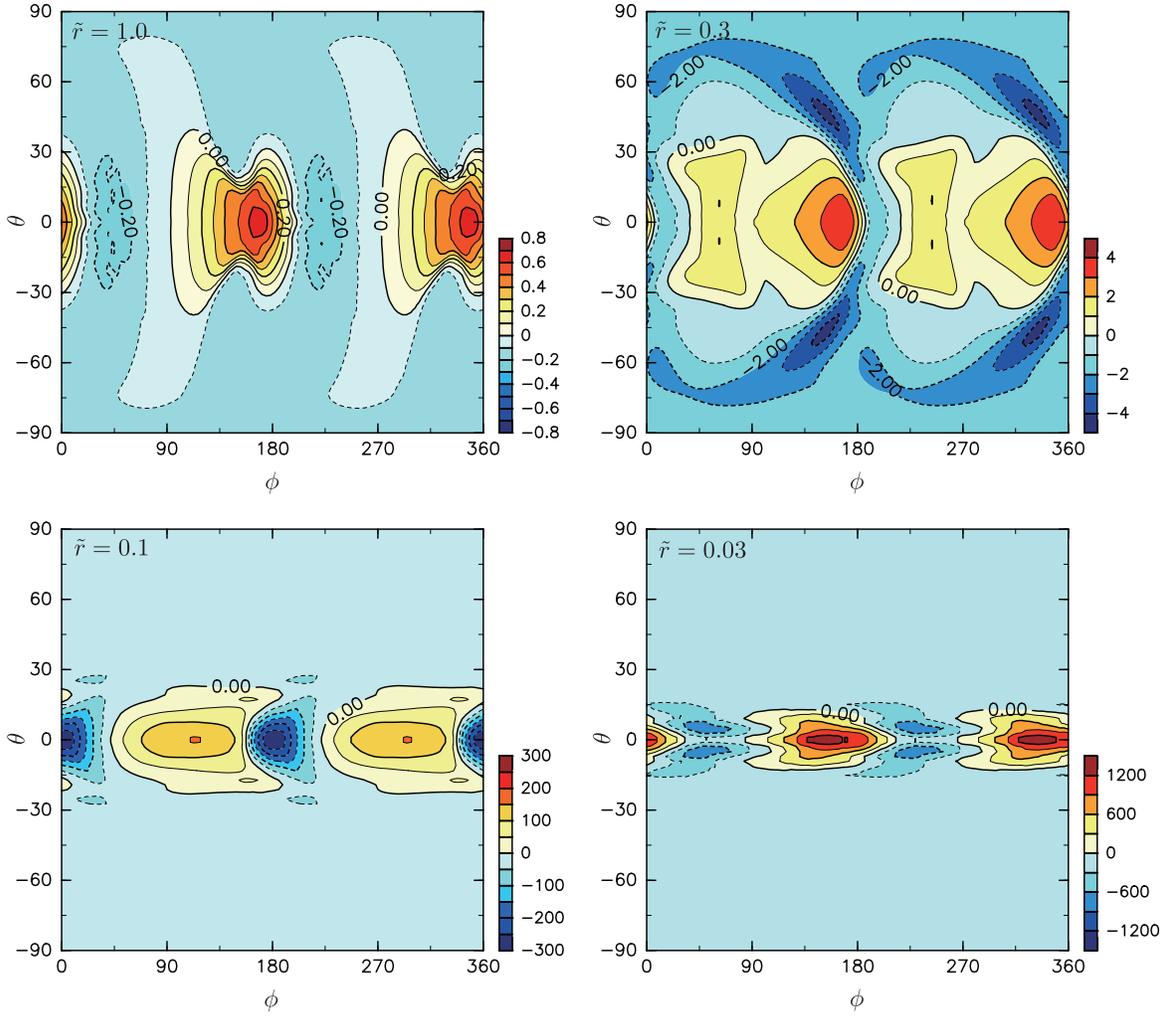}
%
\caption{Mass flux on spheres with radii $\tilde{r} =$ 1.0, 0.3, 0.1,
and 0.03.  Horizontal axis is azimuth angle and vertical axis is
elevation angle on the sphere.  Outward flux is defined to be positive.
$(\phi, \theta) = (0,0)$ and $(180,0)$ correspond to the sub-solar and
anti-solar points, respectively.
\label{Fig_flux_at_spheres}}
\end{figure}

\begin{figure}
\plottwo{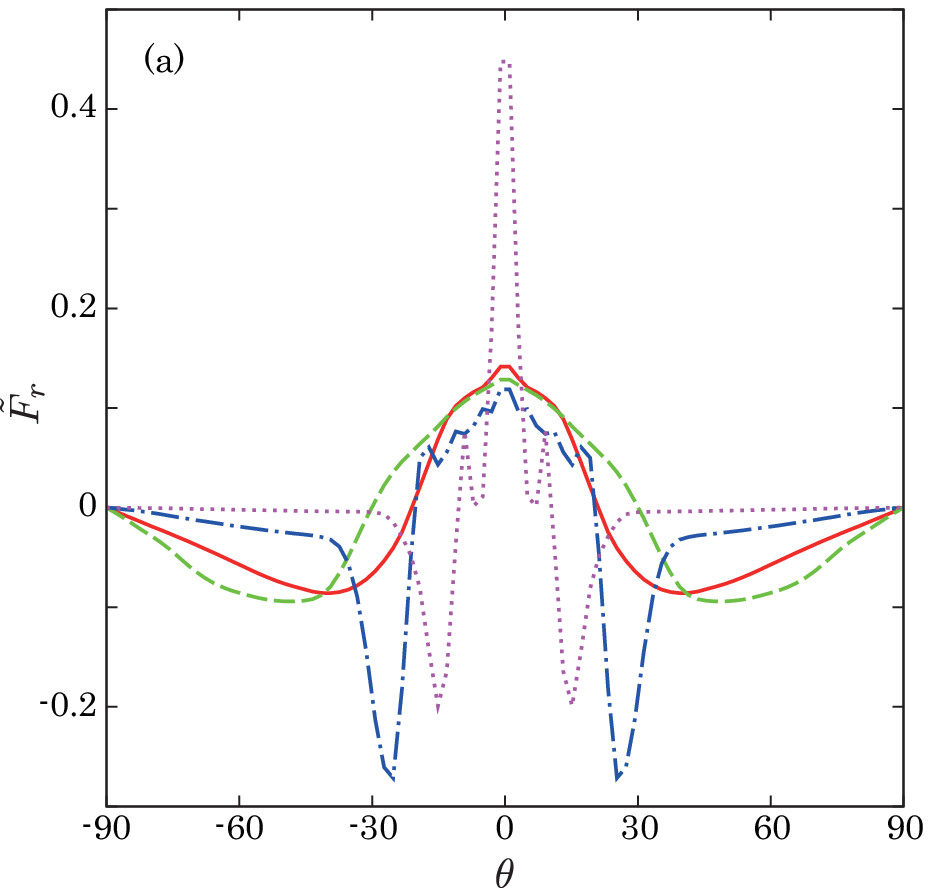}{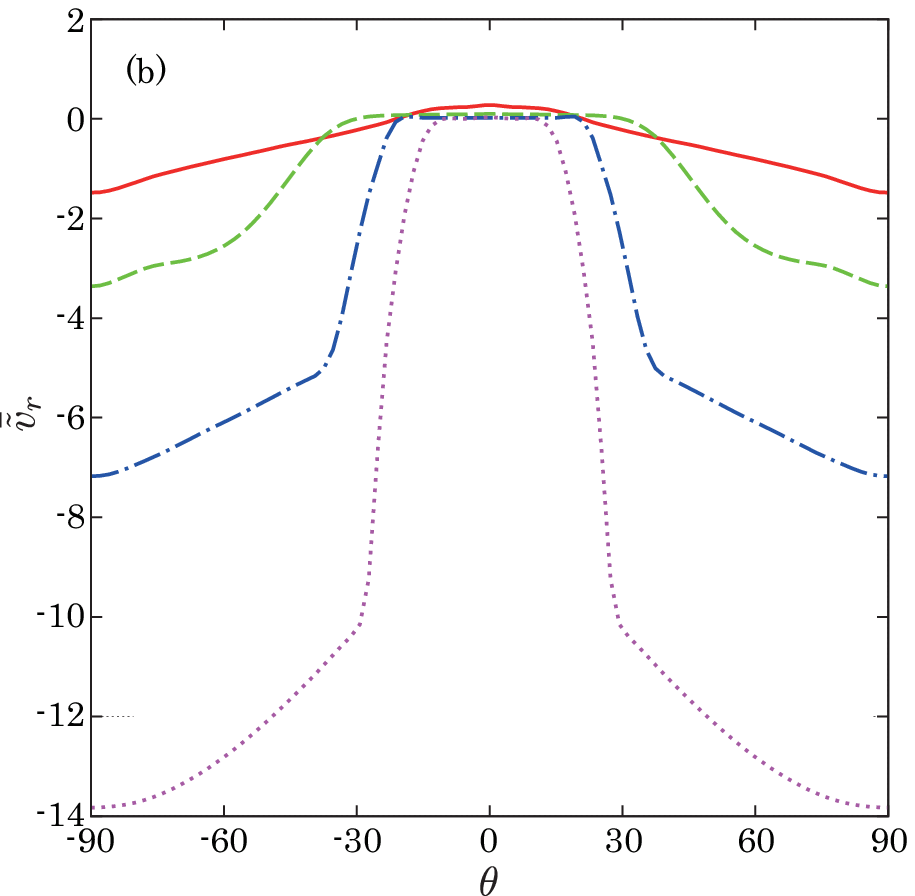}
%
\caption{Left panel shows azimuthally integrated mass flux at the radii
 $\tilde{r} =$ 1.0 (solid red line), 0.3 (dashed green line), 0.1
 (dot-dashed blue line), 0.03 (dotted purple line) as a function of
 elevation angle $\theta$.  Right panel shows azimuthally averaged
 radial velocity at the same radii.
\label{Fig_flux_vs_theta}}
\end{figure}

\begin{figure}
\epsscale{0.5}
\plotone{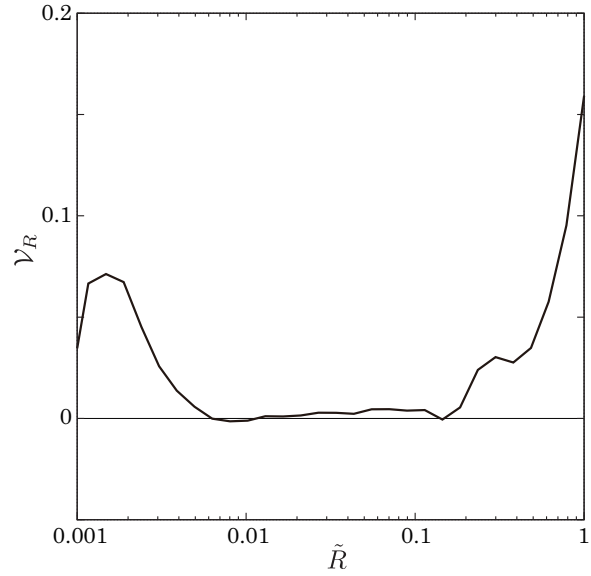}
%
\caption{Azimuthally averaged radial velocity in the midplane normalized
 by local Keplerian velocity ${\cal V}_R$, which corresponds to angle
 between the velocity vector of the flow and the circular orbit at the
 point.
\label{Fig_vr-per-vKep}}
\end{figure}

\begin{figure}
\epsscale{0.5}
\plotone{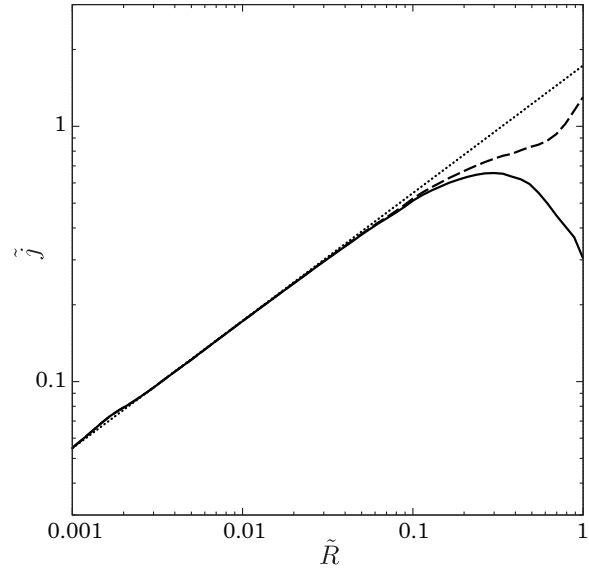}
%
\caption{Azimuthally averaged specific angular momentum at the midplane
 of the circumplanetary disk.  Solid line shows specific angular
 momentum measured in the rotating frame of the Hill coordinate, dashed
 line is the case when it is measured in the inertial frame, and dotted
 line shows specific angular momentum of Keplerian rotation around the
 planet.
\label{Fig_AM_vs_r}}
\end{figure}

\begin{figure}
\epsscale{0.9}
\plottwo{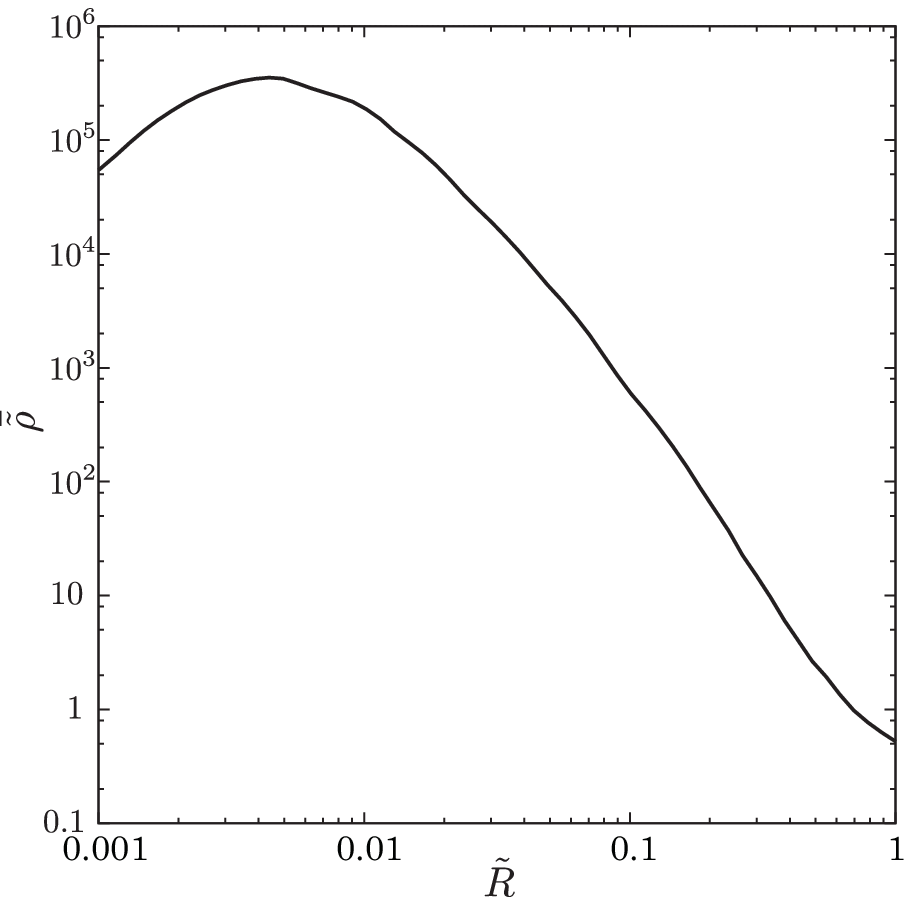}{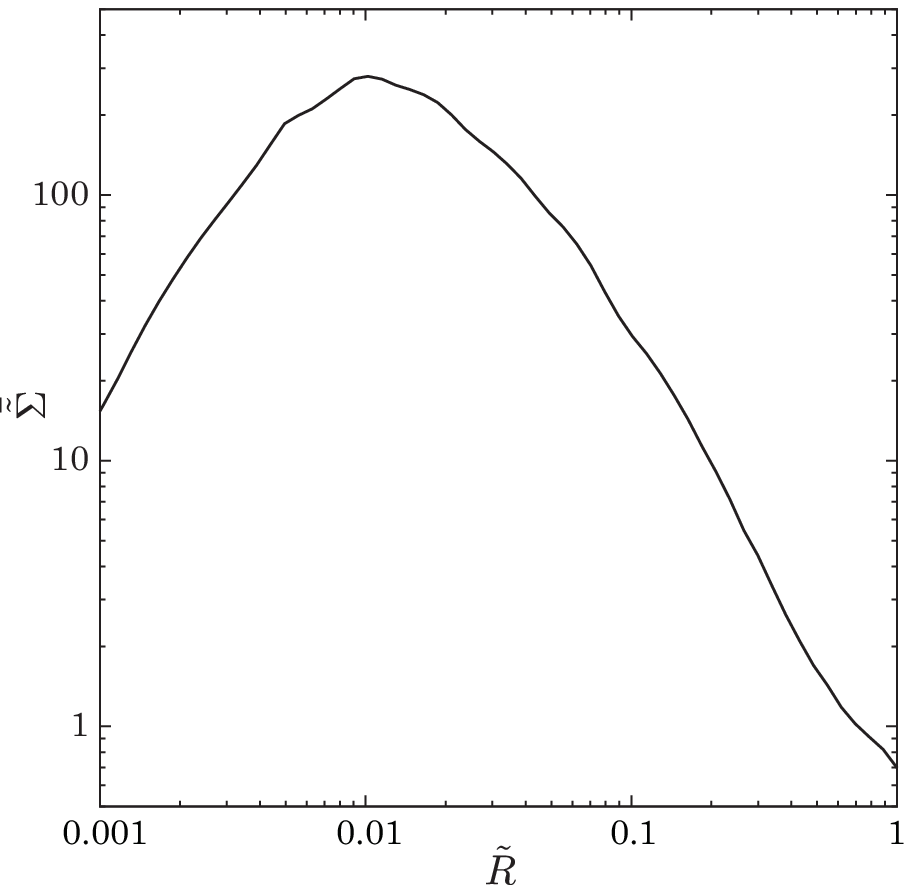}
%
\caption{Azimuthally-averaged density at the midplane (left).  
Azimuthally averaged surface density (right), both as a function of the
 radial distance in the midplane.
\label{Fig_rho_and_sigma_ave}}
\end{figure}

\begin{figure}
\epsscale{1.0}
\plotone{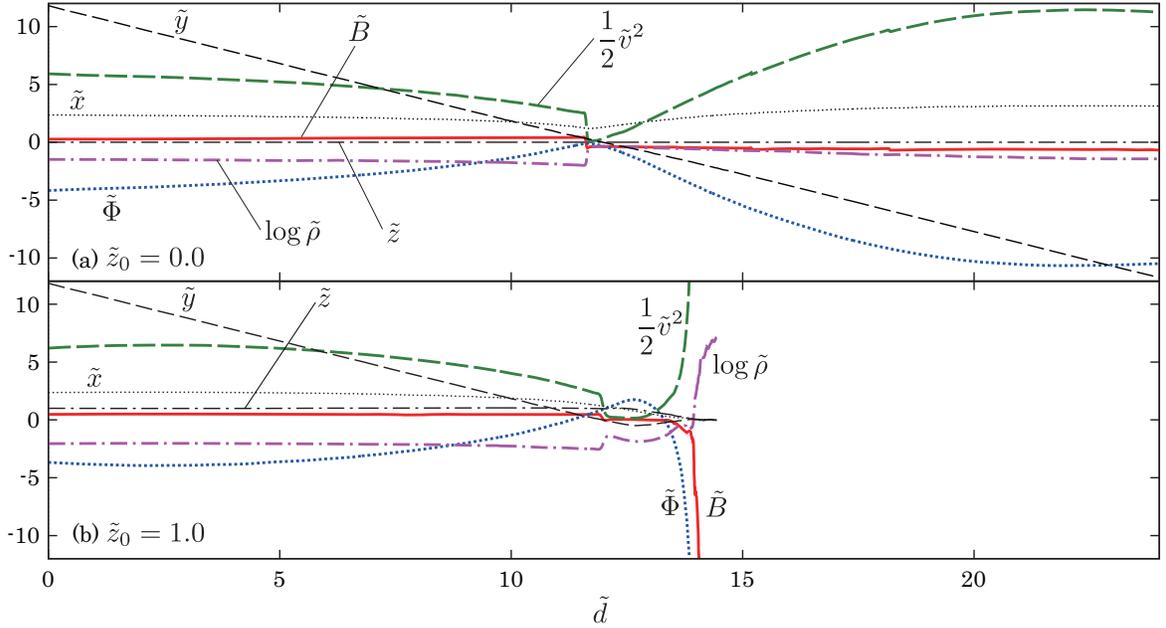}
%
\caption{Quantities along two streamlines starting from two different
 heights.  Horizontal axis $\tilde{d}$ is distance along each streamline
 from the starting point.  Starting points for upper and lower panel are
 $(2.37, \tilde{L}_y/2, 0.0)$ and $(2.37, \tilde{L}_y/2, 1.0)$,
 respectively.  Thick solid (red) line shows Bernoulli integral, thick
 dashed (green) line shows kinetic energy, thick dotted (blue) line
 shows potential energy, and thick dot-dashed line (purple) shows
 logarithm of density.  Thin dotted, dashed, dot-dashed lines show
 $\tilde{x}$, $\tilde{y}$, $\tilde{z}$, respectively.  The streamline
 shown in the lower panel is accreted onto the circumplanetary disk,
 while that in the upper panel is not.  Note that the streamline in the
 case of $\tilde{z}_0 = 1.0$ is truncated at a point where the gas
 element starts rotating around the planet in the circumplanetary disk.
\label{Fig_Bernoulli_wide}}
\end{figure}

\begin{figure}
\epsscale{0.9}
\plotone{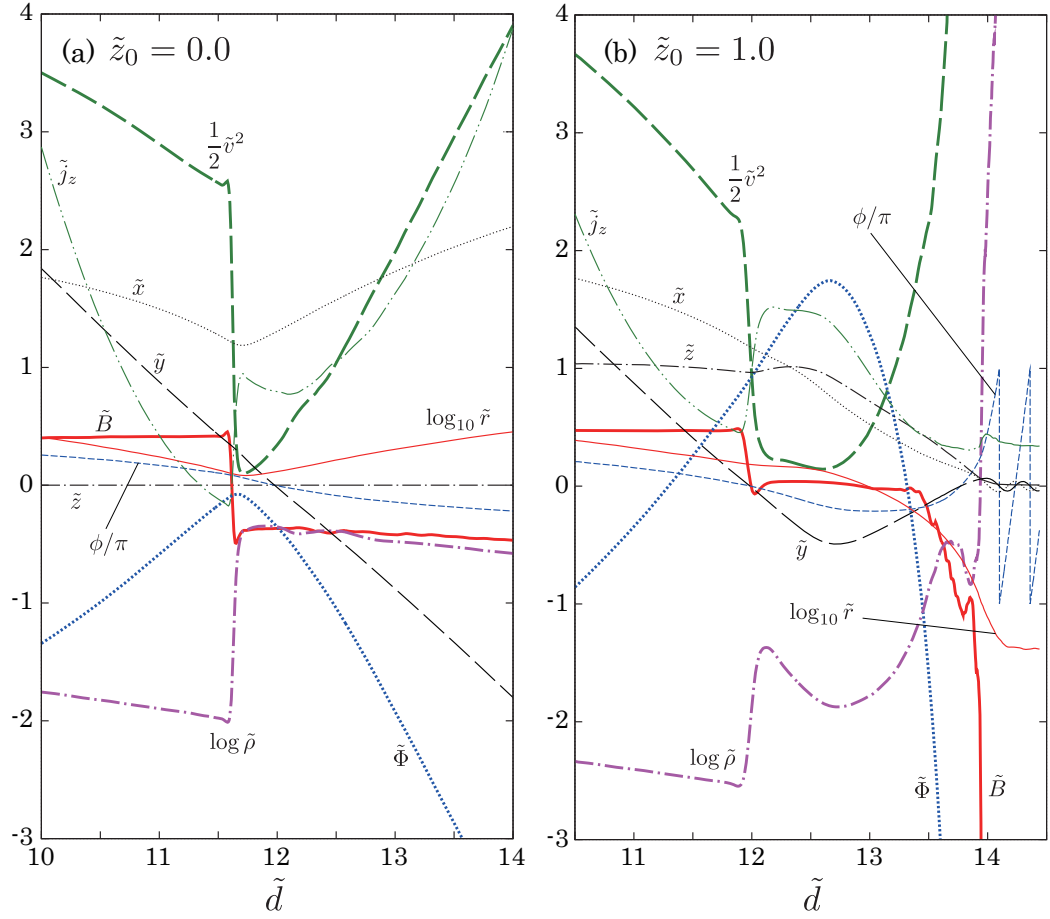}
%
\caption{Close up view of Fig.~\ref{Fig_Bernoulli_wide}.  In addition to
 the quantities shown in Fig.~\ref{Fig_Bernoulli_wide}, azimuth angle
 $\phi$ divided by $\pi$ (thin dashed blue), specific angular momentum
 (thin dot-dot-dashed green), and logarithm of $\tilde{r}$ (thin solid
 red) are also shown.
\label{Fig_Bernoulli_closeup}}
\end{figure}

\begin{figure}
\epsscale{1.0}
\plotone{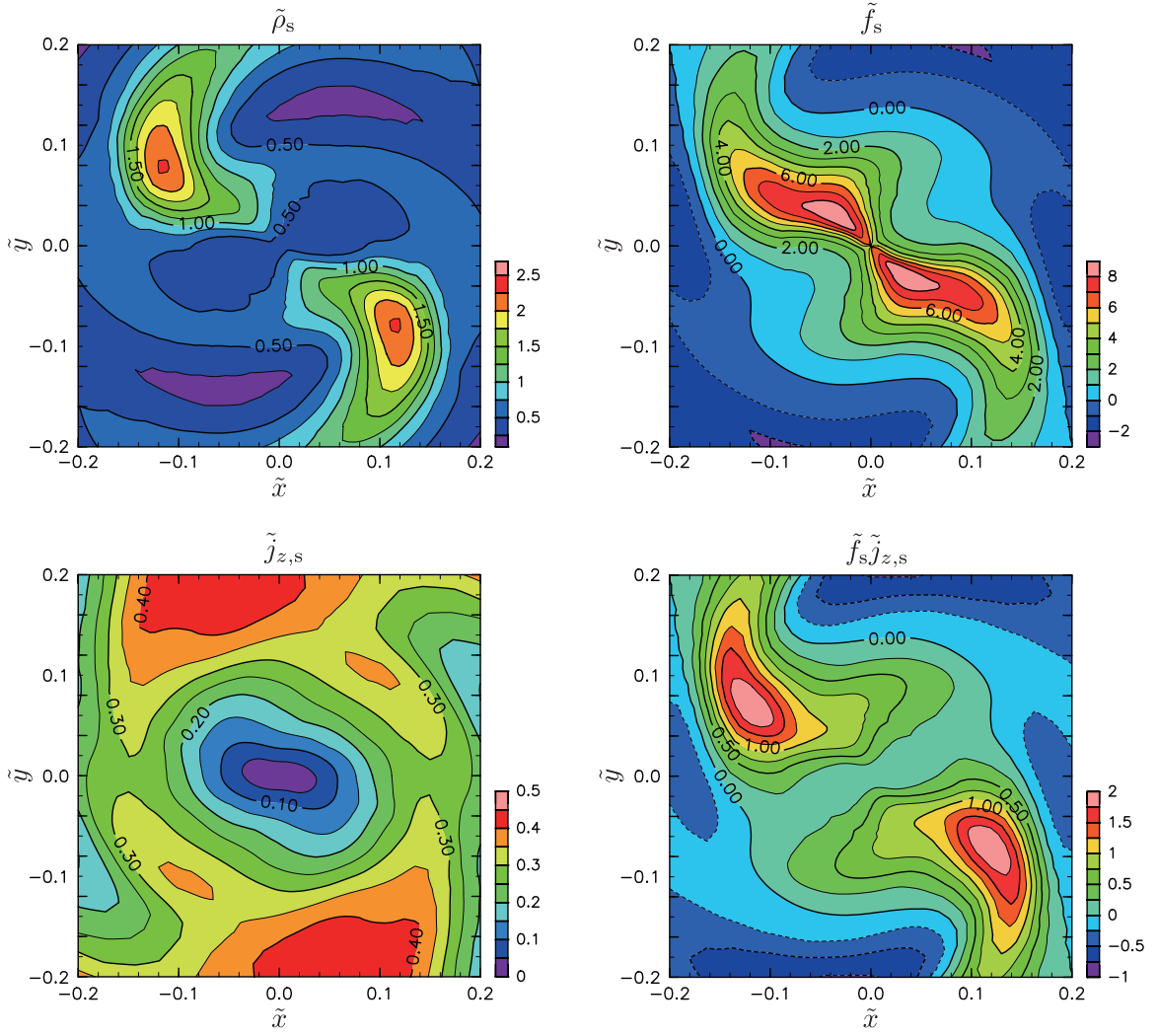}
%
\caption{Distribution of the quantities associated with the accretion
flow at the surface defined by $\tilde{z} = \tilde{z}_{\rm
s}(\tilde{R})$.  Density $\tilde{\rho}_{\rm s}$, mass flux through the
disk surface $\tilde{f}_{\rm s}$, specific angular momentum
$\tilde{j}_{z,\rm s}$, and angular momentum flux through the surface
$\tilde{f}_{\rm s}\tilde{j}_{z,\rm s}$.
\label{Fig_flux_disk-surface}}
\end{figure}

\begin{figure}
\epsscale{0.5}
\plotone{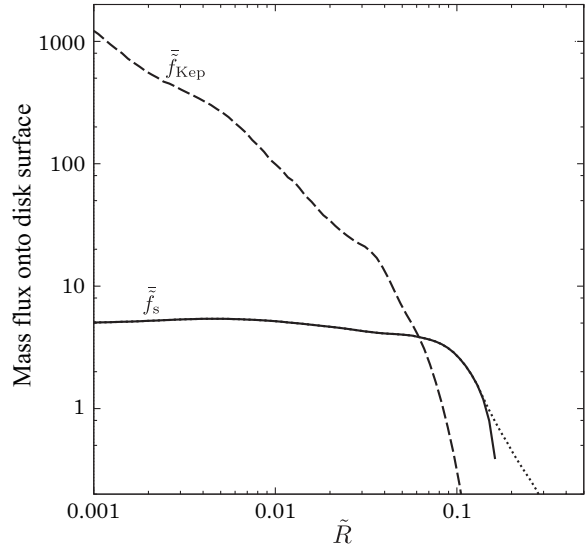}
%
\caption{Azimuthally averaged mass fluxes through the surface $\tilde{z}
 = \tilde{z}_{\rm s}(\tilde{R})$ as a function of radius.  Solid line
 shows $\bar{\tilde{f}}_{\rm s}$ and dashed line shows
 $\bar{\tilde{f}}_{\rm Kep}$.  Dotted line shows $\bar{\tilde{f}}_{\rm
 s}$ but excluding the region where $\tilde{f}_{\rm s}<0$ from the
 integration Eq.~(\ref{f_s_bar}).  $\bar{\tilde{f}}_{\rm s}$ is a flux
 directly observed at the surface, while $\bar{\tilde{f}}_{\rm Kep}$ is
 an effective flux assuming redistribution of radial location where
 specific angular momentum matches that of the local Keplerian rotation.
 (see Eqs.~(\ref{f_s_bar}) and (\ref{f_Kep_bar})).
\label{Fig_f_s_f_Kep}}
\end{figure}

\begin{figure}
\epsscale{0.5}
\plotone{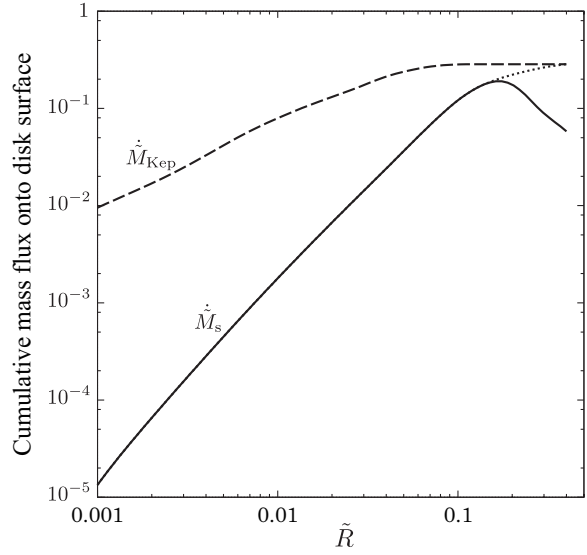}
%
\caption{Mass accretion rate onto the disk surface within a circle of
 radius.  Solid line shows $\dot{\tilde{M}}_{\rm s}$
 (Eq.~(\ref{M_s_dot})) and dashed line shows $\dot{\tilde{M}}_{\rm Kep}$
 (Eq.~(\ref{M_Kep_dot})).  Dotted line shows $\dot{\tilde{M}}_{\rm s}$
 but excluding the region where $\tilde{f}_{\rm s}<0$ from the
 integration of Eq.~(\ref{M_s_dot}).
\label{Fig_M_s_M_Kep}}
\end{figure}

\begin{figure}
\epsscale{0.5}
\plotone{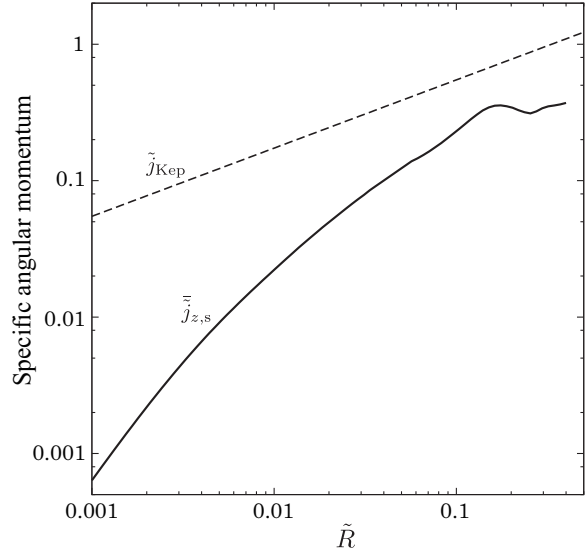}
%
\caption{Azimuthally-averaged specific angular momentum of the accretion
 flow onto circumplanetary disks denoted by $\bar{\tilde{j}}_{z,\rm s}$
 in the text (solid line).  Dashed line is $\tilde{j}_{\rm Kep} =
 \sqrt{3\tilde{r}_{\rm H}^3 \tilde{R}}$, which is specific angular
 momentum for Keplerian rotation around the planet.
\label{Fig_j_zs}}
\end{figure}

\begin{figure}
\epsscale{0.5}
\plotone{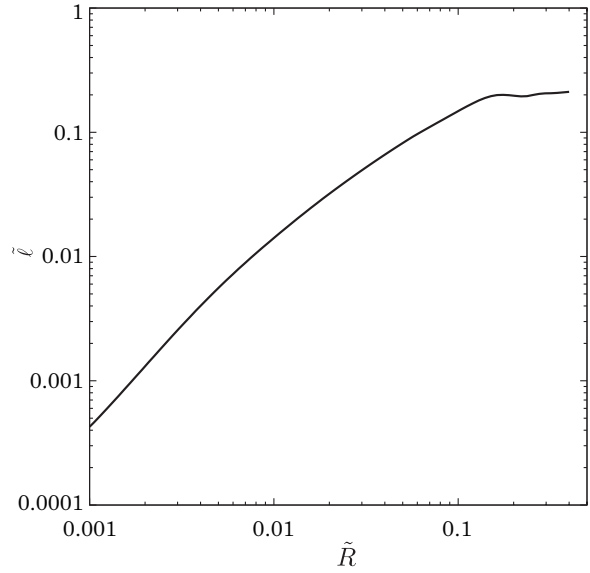}
%
\caption{Mean specific angular momentum of the gas accreting onto the
 part of the circumplanetary disk within the radial distance $\tilde{R}$.
\label{Fig_ell}}
\end{figure}

\begin{figure}
\epsscale{1.0}
\plotone{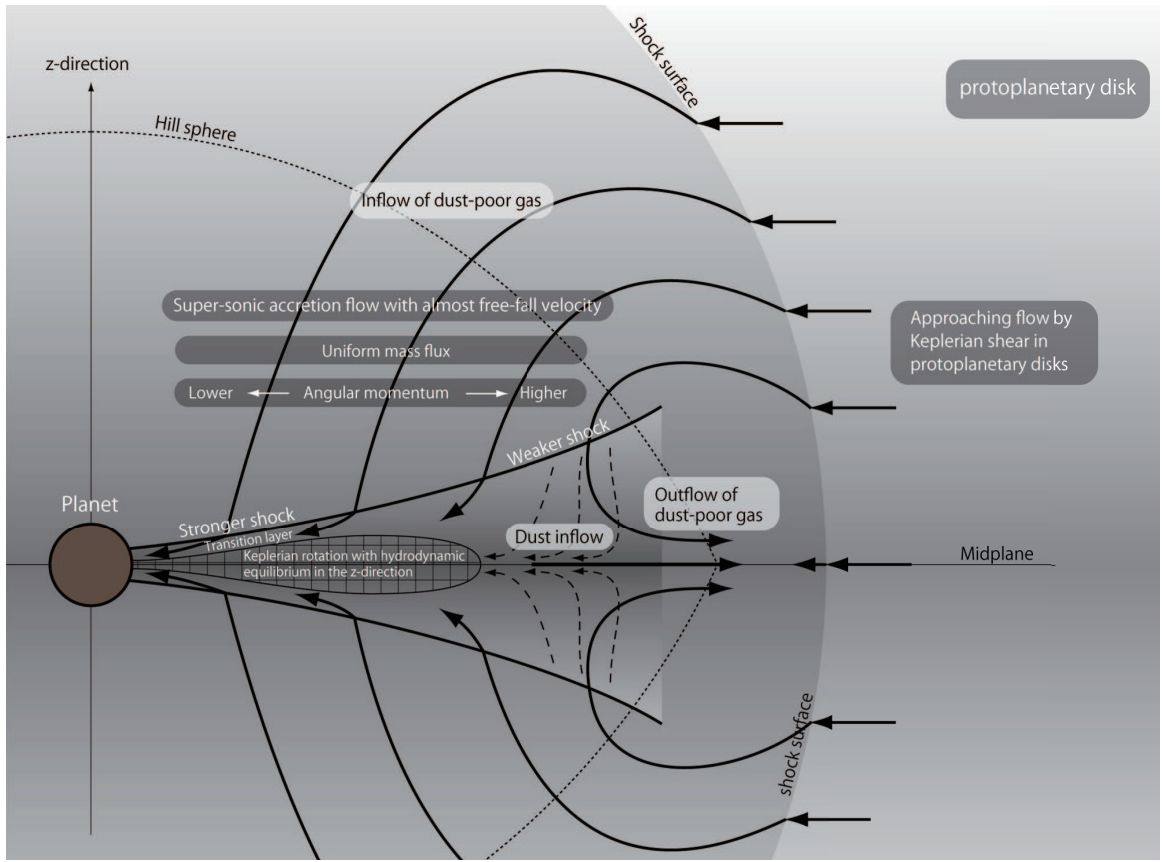}
\caption{Schematic picture of flow structure of circumplanetary disks.
\label{Fig_CP-disk}}
\end{figure}

Our results demonstrate that gas accretion toward and within
circumplanetary disks has a complicated vertical structure.  The width
of the accretion band toward a circumplanetary disk depends on the
initial height of gas elements (Fig.~\ref{Fig_fate}).  Also, after
accretion onto the disk, the gas just below the shock surface migrates
inward, while the gas near the midplane moves radially outward
(Fig.~\ref{Fig_CP-disk}).  Such vertical heterogeneity of the flow may
have significant influence on the dynamical evolution of solid bodies in
the circumplanetary disk, which we will examine in our future work.
We also need to address the effect of viscosity and local calculation,
which would affect the flow quantitatively.



\acknowledgments
We are grateful to Hidekazu Tanaka, Hiroshi Kobayashi, and Satoshi
Okuzumi for valuable comments.  T. T. also thanks Sei-ichiro Watanabe
and Ryuji Morishima for continuous encouragement to advance this.
This work was supported by Center for Planetary Science running under
the auspices of the MEXT Global COE Program entitled ``Foundation of
International Center for Planetary Science''.
We are also grateful for the support by JSPS and NASA's Origins of Solar
Systems Program.
Numerical calculations were carried out on NEC SX-9 at Center for
Computational Astrophysics, CfCA, of National Astronomical Observatory
of Japan.
A part of the figures were produced by GFD-DENNOU Library.

\clearpage





%
%
%

%
%
%

\end{document}